\documentclass[pra,superscriptaddress,twocolumn,floatfig,showpacs,floatfix]{revtex4-1}
\usepackage{graphicx,color}
\usepackage{amsmath,amssymb,bm}
\usepackage{braket}		
\usepackage{multirow}

\definecolor{darkred}{rgb}{0.90,0,0}
\definecolor{darkgreen}{rgb}{0,0.60,.2}
\definecolor{darkblue}{rgb}{0,0,1}
\definecolor{grey}{cmyk}{0,0,0,0.25}
\definecolor{orange}{cmyk}{0,0.6,1,0}

\begin{document}
\title{Properties of the single-site reduced density matrix in the Bose-Bose resonance model in the ground state and in quantum quenches}

\author{F. Dorfner}
\affiliation{Department of Physics and Arnold Sommerfeld Center for Theoretical Physics,
Ludwig-Maximilians-Universit\"at M\"unchen, D-80333 M\"unchen, Germany}
\author{F. Heidrich-Meisner}
\affiliation{Department of Physics and Arnold Sommerfeld Center for Theoretical Physics,
Ludwig-Maximilians-Universit\"at M\"unchen, D-80333 M\"unchen, Germany}

\begin{abstract}
We study properties of the single-site reduced density matrix in the Bose-Bose resonance model  as a function of system parameters.
This model describes a single-component Bose gas with a resonant coupling to a diatomic molecular state, here defined on a lattice.
A main goal is to demonstrate that the eigenstates of the single-site reduced density matrix have structures that are characteristic for the various 
quantum phases of this system. Since the Hamiltonian conserves only the global particle number but not the number of bosons and 
molecules individually, these eigenstates, referred to as optimal modes,  can be nontrivial linear combinations of bare eigenstates of the molecular and boson particle number.
We numerically analyze the  optimal modes  and their weights, the latter giving the  importance of the corresponding state, in 
the ground state of the Bose-Bose resonance model. 
We find that the single-site von Neumann entropy is sensitive to the location of the phase boundaries.
We explain the structure of the optimal modes and their weight spectra using perturbation theory and via a comparison to results for the single-component Bose-Hubbard model.
We further study the dynamical evolution of the optimal modes and of the single-site entanglement entropy in two quantum quenches that cross phase boundaries of the model
and show that these quantities are thermal in the steady state.
For our numerical calculations, we use the density matrix renormalization group method for ground-state calculations and time evolution in a Krylov subspace for the quench dynamics as well as exact diagonalization.
\end{abstract}

\maketitle



\section{Introduction}
\label{sec:introduction}

Studying  entanglement measures in the vicinity of quantum phase
transitions of many-body Hamiltonians has become a very active
field and is quite a useful tool. For the characterization of quantum phases
the finite-size scaling properties of entanglement entropies such as the von Neumann entropy
have turned out to be very informative, providing information about gapped phases
through area laws and the number of gapless modes via the central charge for
gapless systems with a conformally invariant low-energy theory \cite{eisert-review,schollwoeck11}.
Moreover, additional information can be extracted from the entanglement spectrum (see, e.g., \cite{haldane,fidkowski10,thomale10,pollmann10,turner10,laeuchli10,alba13}).

Numerical work has suggested that the single-site entanglement entropy
can be sensitive to quantum phase transitions in interacting fermionic systems in one
dimension such as the extended Hubbard model \cite{gu}. This has further been explored and applied to
spin systems \cite{legeza06} as well as to the Bose-Hubbard model (BHM) \cite{giorda04}.
The single-site entanglement entropy $S_{\rm vN}^{(1)}$ is directly linked to the eigenvalues $w_\alpha$
of the single-site reduced density matrix $\rho^{(1)}$ obtained by tracing out the degrees of freedom of all sites but one (defining the environment $E$) from the ground-state
wave function:
\begin{gather}
\rho^{(1)} = \mbox{tr}_{E} (| \psi \rangle \langle \psi |) = \sum_{\alpha} w_{\alpha}| \alpha \rangle \langle \alpha | \\
S_{\rm vN}^{(1)} = - \sum\limits_{\alpha} w_{\alpha} \ln(w_{\alpha}).
\end{gather}
Here, $| \alpha \rangle$ are the eigenstates of $\rho^{(1)}$ and the $w_{\alpha}$ are their weights.
Since the individual  particle numbers for spin up and down are conserved in the Hubbard model  and since the local
Hilbert space is only two- or four-dimensional for spin-1/2 and Hubbard models, respectively,
one can easily see that there are very  few free parameters, taking into account also normalization
$\sum_{\alpha} w_{\alpha} = 1$.
Therefore, for a spin-1/2 system with spin-inversion symmetry, $S_{\rm vN}^{(1)} = \mbox{ln}(2)$, independent of system size
and the actual model. For the Fermi-Hubbard model at half filling and vanishing magnetization, there is only one free parameter \cite{gu}.

We will be interested in systems with large local {\it bosonic} Hilbert spaces, where some crucial differences
arise.  First, the
local Hilbert space is much larger and second, in models that do not preserve particle number, the eigenstates  $| \alpha \rangle$
of $\rho^{(1)}$ do not need to be eigenstates of the local particle number (plus possible additional $U(1)$ symmetries).
This is most notably the case for systems with phononic degrees of freedom such as the Holstein model \cite{holstein}.
Originally intended as a means to improve numerical methods, Zhang, Jeckelmann and White~\cite{zhang98} introduced the term
optimal modes for the eigenstates of the single-site reduced density matrix. 
Their idea was to set up algorithms in an effective Hilbert space obtained by truncating in the spectrum of the single-site reduced
density matrix. This gives a computational advantage whenever the weight spectrum decays sufficiently fast.
This concept has been used in
exact diagonalization studies \cite{zhang99,weisse00,alvermann10} and density matrix renormalization group algorithms \cite{friedman00,guo12,bruognolo14,friend15,brockt15}  but also bears useful
information about the equilibrium \cite{zhang98} and non-equilibrium physics \cite{dorfner15} of such systems.
In general, for systems with bosonic degrees of freedom,  $\rho^{(1)}$ and hence also the single-site entanglement
entropy can therefore harbor much more information than in fermionic or spin systems.

In our work we consider a bosonic model with a global $U(1)$ symmetry yet two species of bosons
(labeled $s=a,m$) whose particle numbers are not individually conserved. This system thus possesses non-trivial optimal modes, which, as a function of model parameters, can undergo a mixing of the contributions coming from the bare eigenstates
of both particle numbers, which one typically uses to set up a convenient basis for numerical methods.

Concretely, we study the so-called one-dimensional (1D)
Bose-Bose-resonance model (BBRM) \cite{romans04,radzihovsky04,sengupta05,radzihovsky08,eckolt10,ejima11,bhaseen12} that describes a Bose gas with
repulsive contact interactions plus resonant interactions, here  defined  on a lattice (for a study in two and three dimensions we refer to \cite{deparny15}).
The resonant interaction mimics the physics of a Feshbach resonance \cite{bloch08,chin10}:
When two atoms meet on the same site, they can form a molecule.
The Hamiltonian reads:
\begin{eqnarray}
H &=& H_{\rm{BH,a}} + H_{\rm BH,m} + H_{\rm{int}} + H_{\rm{F}} + H_{\rm{D}}  \label{hamiltonian} \\
H_{{\rm BH},{\rm s}} &=& -t_{\rm s} \sum\limits_{j} ({s}_{j}^{\dagger} {s}_{j+1} + h.c.)  
 + \frac{U_{\rm s}}{2} \sum\limits_j n_{j,\rm s}(n_{j,\rm s} - 1) \nonumber  \\
H_{\rm{int}} &=& U_{\rm{a,m}} \sum\limits_{j} n_{j,\rm a}n_{j,\rm m}  \nonumber\\
H_{\rm{F}} &=& g \sum\limits_{j} ({\rm m}_{j}^{\dagger} {\rm a}_{j} {\rm a}_{j} + {\rm m}_{j} {\rm a}_{j}^{\dagger} {\rm a}_{j}^{\dagger})\nonumber  \\
H_{\rm{D}} &=& \epsilon_{\rm m} \sum\limits_j n_{j,\rm m}. \nonumber
\end{eqnarray}
The operator ${s}_{j}$ (${s}_{j}^{\dagger}$) annihilates (creates) a boson of species ${s}={a,m}$ and $n_{j,\rm s} = {\rm s}_{j}^{\dagger} {\rm s}_{j}$ measures the particle density of species $s$ at site $j$.
The full Hamiltonian $H$ consists of five parts:
A Bose-Hubbard term $H_{\rm{BH,s}}$ for each species, a repulsive on-site inter-species interaction term $H_{\rm{int}}$,
the Feshbach-coupling term $H_{\rm F}$ and the detuning term $H_{\rm D}$.
The Feshbach-coupling term describes the conversion of two atoms into a molecule and vice versa.
Because of this conversion the Hamiltonian conserves only the total number of particles $N_{\rm T} = N_{\rm a} + 2 N_{\rm m}$ where
$N_{\rm s} = \sum_j \langle n_{j,\rm s}\rangle$ denotes the particle number of the individual species $s$.

\begin{figure}[t]
\includegraphics[width=.80\columnwidth]{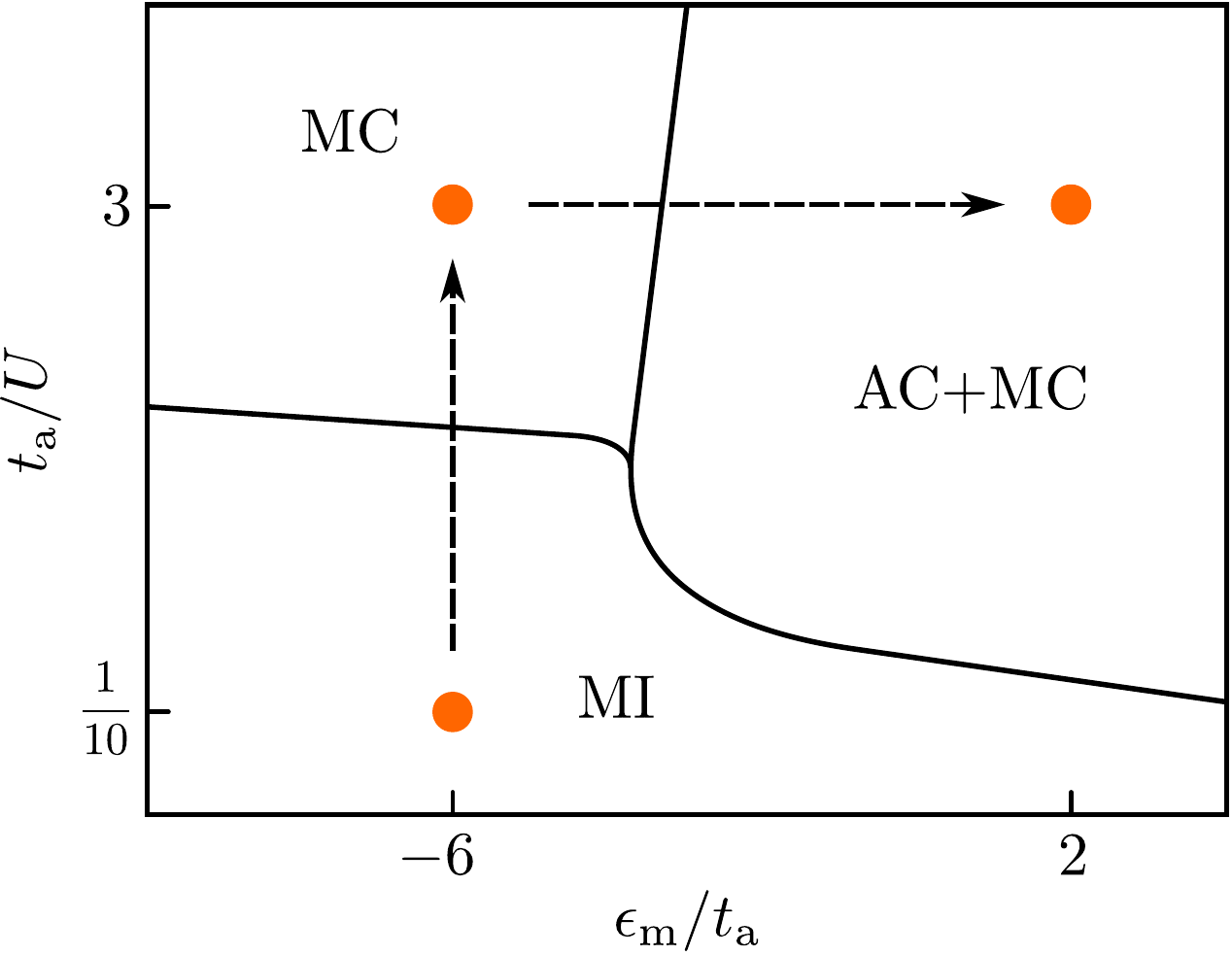}
\caption{(Color online)
Schematic phase diagram of the BBRM model based on the results of Ref.~\cite{ejima11}. The labels denote the three phases:
molecular condensate (MC), molecular and atomic condensate (AC+MC) and Mott-insulator (MI). The horizontal and vertical dashed lines mark the two 
trajectories through the phase diagram along which we compute entanglement properties and optimal modes, both in equilibrium and in quantum quenches.}
\label{fig:phase-diag}
\end{figure}

Our main goal is to elucidate  the behavior of the single-site entanglement entropy and the structure of optimal modes in
the quantum phases of this model and, in particular, in the vicinity of the phase transitions.
Moreover, we consider quantum quenches between different phases and investigate the changes in the optimal
modes in non-equilibrium dynamics. 

Throughout this work, we focus on the following set of parameters, for which the phase diagram of the model is known from Ref.~\cite{ejima11}: $t_{\rm m}=t_{\rm a}/2$, $U_{\rm a}/2=U_{\rm m}/2=U_{\rm am}=g=U$.
Moreover, we work at fixed filling $N_{\rm T}/L=2$, where $L$ is the number of sites and we set $t_{\rm a}=1$.
The phase diagram is schematically shown in Fig.~\ref{fig:phase-diag}. There are three phases \cite{ejima11}: a Mott-insulator (MI), a molecular condensate (MC) and a phase in which both atoms and molecules quasi-condense (AC+MC).
Note that by using the term condensate we follow the terminology of \cite{ejima11}, yet of course, in one dimension, there can only be quasi-long range order.

Our main results are:
First, we find that the first derivative of the local von Neumann entropy with respect to the detuning displays a maximum at the position of the phase boundary between the MC and AC+MC phases.
Furthermore, we observe that the single-site entanglement entropy is a monotonically increasing (decreasing) function of system size in the MI (MC) phase.
We provide qualitative arguments for this behavior and contrast it to the Mott-insulator-to-superfluid transition in the Bose-Hubbard model.
This change in the system-size dependence leads to features in the vicinity of the phase transition as our numerical data suggest.
Next, we study the weights and optimal-mode spectra as a function of model parameters and explain their behavior in the three phases indicated by the points in Fig.~\ref{fig:phase-diag} by using density matrix renormalization group (DMRG) \cite{white92,schollwoeck05} simulations of the BBRM, perturbation theory and a comparison to numerical results for the Bose-Hubbard model.

While our main goal is the investigation of entanglement properties and of optimal-mode structures,
our work is also one of the first numerical studies of quantum quenches in interacting Bose gases
with {\it resonant} interactions. This is relevant to ultra-cold quantum gas physics \cite{bloch08} where such quenches 
play an important role and were studied experimentally in Bose gases (see, e.g., \cite{claussen2002,donley2002}). Recent experiments have investigated quenches in Bose gases to 
unitarity \cite{makotyn2014}.
Here, however, we do not aim at making contact  with any  experiment with a Feshbach resonance.
For recent studies of non-equilibrium properties in fermionic and bosonic systems coupled to bound states via Feshbach interactions, see \cite{timmermans99,holland01,thalhammer06,scott12,breid13,yin16}.
We here report  results for two quantum quenches where we start from a point deep in the MI (MC) phase and quench the system over to the MC (AC+MC) phase (Fig.~\ref{fig:phase-diag}).
The pre- and postquench parameters are chosen such that they correspond to the cases studied in the ground-state section, i.e., those indicated in Fig.~\ref{fig:phase-diag}, deep in the respective phases.
In addition to the  local von Neumann entropy  and the structure of the optimal modes (Sec.~\ref{sec:quenchstrucom}), we discuss the time evolution of the momentum distribution function for both, molecules and atoms.
Because of the large bosonic Hilbert space and fluctuations, the question of thermalization is already interesting for a single-site object.
We find that the single-site reduced density matrix is thermal in the steady state by comparison to the corresponding expectation values in the canonical ensemble.

The plan of the paper is the following.
We start by defining the computational basis, the relevant observables and numerical methods in Sec.~\ref{sec:obs}.
In Sec.~\ref{sec:gsprops}, we study the ground-state properties of our system.
More specifically, the number of molecules and the single-site von Neumann entropy (Sec.~\ref{sec:gs_nm_svn}) as well as the weight spectrum and structure of the optimal modes (Sec.~\ref{sec:gs_w_om}) are studied as a function of detuning and inverse interaction strength along two different trajectories through the phase diagram.
Section~\ref{sec:ququenches} illustrates the behavior of the BBRM system under a global quench from the MI to the MC phase and from the MC to the AC+MC phase.
We conclude in Sec.~\ref{sec:conclusion} with a summary of our results.


\section{Observables, definitions and numerical methods}
\label{sec:obs}

For the BBRM we measure the atomic and molecular particle numbers, the momentum distribution function, the optimal-mode weights and spectra (the optimal mode expressed in the bare occupation number basis) as well as the single-site entanglement entropy for the ground state and during quenches between different phases.

\subsection{Optimal modes and von Neumann entropy}
\label{sec:optmodes}

The optimal modes and their weights can be obtained by diagonalizing the single-site reduced density matrix
\begin{eqnarray}
\rho^{(1)} &=& \text{tr}_{\rm E} \rho \notag \\
&=& \sum\limits_{n,n'} \sum\limits_j \psi_{nj}^* \psi_{n'j} \ket{n}\bra{n'} \label{eq:rho} \\
&=& \sum\limits_{\alpha} w_{\alpha} \ket{\alpha}\bra{\alpha}, \notag \\
| \alpha \rangle &=& \sum\limits_{n} \langle n | \alpha \rangle | n \rangle \notag
\end{eqnarray}
where
$\rho$ is the density matrix of the full system, the $| \alpha \rangle$ denote the optimal modes and $w_{\alpha}$ denote their weights (relative importance).
We refer to the decomposition coefficients $|\langle \alpha | n \rangle|^2$ as optimal-mode spectrum.
From the weights one can directly calculate the single-site von Neumann entropy
\begin{equation}
S_{\rm{vN}}^{(1)} = - \sum\limits_{\alpha} w_{\alpha} \log(w_{\alpha}).
\end{equation}

\subsection{Computational basis}
\label{sec:basis}
To numerically simulate the BBRM model we need a two-component basis consisting of both an atomic and molecular part.
As already stated, because of the Fesh\-bach term in Eq.~\eqref{hamiltonian} they do not completely decouple and only the total particle number $N_{\rm T}$ is conserved.
As a consequence, the basis splits into blocks of tensorproducts between atomic and molecular subbasis sets with a fixed number of particles $N_{\rm T}^{(1)}$.
For bare local states we use the convention
\begin{equation}
\ket{n} = \ket{N_{\rm a}^{(1)};N_{\rm m}^{(1)}},
\end{equation}
where $N_{\rm a}^{(1)}$ ($N_{\rm m}^{(1)}$) denotes the local particle number of atoms (molecules) leading to the total local particle number of $N_{\rm T}^{(1)} 
= N_{\rm a}^{(1)}+2 N_{\rm m}^{(1)}$.
\begin{table}
\centering
\begin{tabular}{c | c c c c c}
$N_{\rm T}^{(1)}$ & \multicolumn{5}{c}{$n$: $\ket{N_{\rm{a}}^{(1)},N_{\rm{m}}^{(1)}}$} \\
\hline
0 & 0: $\ket{0;0}$ & & & & \\
1 & 1: $\ket{1;0}$ & & & & \\
2 & 2: $\ket{2;0}$ & 3: $\ket{0;1}$ & & & \\
3 & 4: $\ket{3;0}$ & 5: $\ket{1;1}$ & & & \\
4 & 6: $\ket{4;0}$ & 7: $\ket{2;1}$ & 8: $\ket{0;2}$ & & \\
5 & 9: $\ket{5;0}$ & 10: $\ket{3;1}$ & 11: $\ket{1;2}$ & & \\
6 & 12: $\ket{6;0}$ & 13: $\ket{4;1}$ & 14: $\ket{2;2}$ & 15: $\ket{0;3}$ & \\
7 & 16: $\ket{7;0}$ & 17: $\ket{5;1}$ & 18: $\ket{3;2}$ & 19: $\ket{1;3}$ & \\
8 & 20: $\ket{8;0}$ & 21: $\ket{6;1}$ & 22: $\ket{4;2}$ & 23: $\ket{2;3}$ & 24: $\ket{0;4}$ \\
\vdots & & & & &
\end{tabular}
\caption{
Bare local basis sets for a fixed number of total particles $N_{\rm T}^{(1)}$ on a site.
The number $n$ left of the respective state is its position in the full local basis.
As $N_{\rm T}^{(1)}$ increases, a growing number of states can mix due to the Feshbach term.
}
\label{tab:localbasis}
\end{table}
Table~\ref{tab:localbasis} shows a subset of the states that make up the local basis labeled by index $n$.
They are ordered first, by the total number of particles on the site $N_{\rm T}^{(1)}$ and  second, by the number of molecules on the site.
Since the Feshbach term can not change $N_{\rm T}^{(1)}$, only states that are in the same row in the table can mix in the local reduced density matrix.
In general, for a fixed $N_{\rm T}^{(1)}$, there are $\lfloor(N_{\rm T}^{(1)}+2)/2\rfloor$ local states that can mix.

\subsection{Momentum distribution function}
The momentum distribution function is obtained from a Fourier transformation of the respective one-particle density matrices of atoms and molecules
\begin{eqnarray}
n_k^{s} &=& \frac{1}{L} \sum\limits_{j,j'} \text{e}^{-ik(j-j')} \braket{s_j^{\dagger} {s}_{j'}} \notag \\
&=& \sum\limits_x \text{e}^{ikx} \braket{\Gamma(x)}, \label{eq:mdf}
\end{eqnarray}
where $\Gamma(x) = \frac{1}{L} \sum\limits_j s_j^{\dagger} s_{j+x}$ is the correlator between site $j$ and $j+x$ for species $s$.

\subsection{Numerical methods}
\label{sec:methods}

We use two wave-function based numerical methods: exact diagonalization (ED) and the density matrix renormalization group (DMRG) method.
DMRG is a matrix-product states based algorithm \cite{white92,schollwoeck05,schollwoeck11}.
To treat the rather large local state-space dimension efficiently we employ the so called single-site DMRG with subspace expansion \cite{hubig15}.
This particular method has the advantage that it scales with $O(d^2)$ instead of $O(d^3)$ as the well-known two-site DMRG where $d$ denotes the size of the local Hilbert space.
To keep the size of the Hilbert space manageable we conserve the number of particles $N_{\rm T}$ during the simulation.

The ED method is used to get numerically exact data for small systems.
Time evolution is performed in the Krylov space with the time step chosen small enough to be accurate keeping $20$ Krylov states.
As mentioned earlier, since both atoms and molecules are bosons the local dimension $d$ is very large.
To deal with this we exploit particle number conservation, translation symmetry and reflection symmetry on a lattice with periodic boundary conditions.


\section{Ground-state properties}
\label{sec:gsprops}

Here, we present our results for properties of the single-site entanglement entropy and of the optimal modes in the three phases, the molecular condensate (MC), atomic and molecular condensate (AC+MC) and the Mott insulating (MI) phase.

\subsection{Single-site von Neumann entropy and molecular density}
\label{sec:gs_nm_svn}

In this section we study the number of molecules and the von Neumann entropy as a function of detuning $\epsilon_{\rm m}$ and interaction strength $U$.
For this purpose we choose two trajectories  through the phase diagram as indicated by the arrows in Fig.~\ref{fig:phase-diag}.

\subsubsection{MC to AC+MC phase}

\begin{figure}[t]
\includegraphics[width=.96\columnwidth]{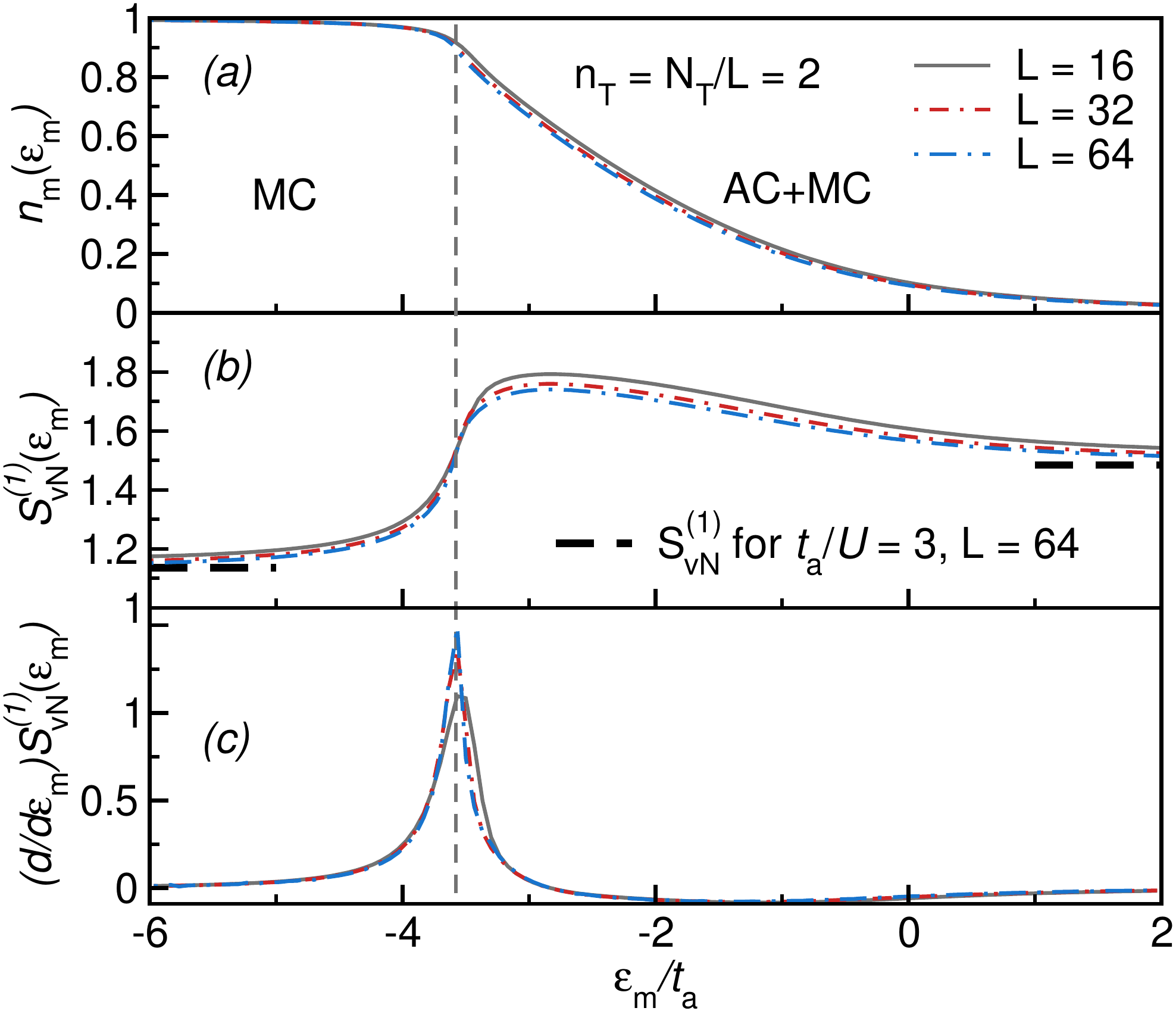}
\caption{(Color online)
(a) Density of molecules and (b) single-site von Neumann entropy for the ground state along the trajectory between the MC ($\epsilon_{\rm m}= -6t_{\rm a}, t_{\rm a}/{U} = 3$) and AC+MC ($\epsilon_{\rm m}= 2t_{\rm a}, t_{\rm a}/{U} = 3$) phases.
The horizontal dashed lines show the values for the local entanglement entropy calculated for a Bose-Hubbard model at unit filling and $U/t_{\rm a}=20$ in the MC limit and double filling and $U/t_{\rm a}=10$ in the AC+MC limit.
(c) First derivative of $S_{\rm vN }^{(1)}$ with respect to $\epsilon_{\rm m}$.
The vertical dashed line marks the position $\epsilon_{\rm m}^c$ of the phase transition taken from \cite{ejima11}.
The data are calculated using DMRG with a local cutoff $N_{\rm T}^{(1)} = 30$ and a bond dimension $D=200$.
}
\label{fig:nm_svn_gs_mc_acmc}
\end{figure}

The first contour connects the MC to the AC+MC phases by varying $\epsilon_{\rm m}$ only.
The results for this case are presented in Fig.~\ref{fig:nm_svn_gs_mc_acmc}.
Figure~\ref{fig:nm_svn_gs_mc_acmc}(a) shows the density of molecules as a function of the detuning.
For  $\epsilon_{\rm m} = -6t_{\rm a}$, there are practically only molecules present in the system.
This is  expected because for a fixed interaction $U/t_a$ the detuning regulates which one of the two species, atoms or molecules, are favored in the ground state, and thus
\begin{equation}
\lim_{\epsilon_{\rm m} \to -\infty} N_{\rm m}(\epsilon_{\rm m}) = \frac{N_{\rm T}}{2}. \label{eq:nmminusinfty}
\end{equation}
For $\epsilon_{\rm m} = 2t_{\rm a}$, the number of molecules is small  since
\begin{equation}
\lim_{\epsilon_{\rm m} \to \infty} N_{\rm m}(\epsilon_{\rm m}) = 0\,. \label{eq:nmplusinfty}
\end{equation}

Figure~\ref{fig:nm_svn_gs_mc_acmc}(b) illustrates the dependence of the single-site von Neumann entropy on the detuning.
For large positive and negative $\epsilon_{\rm m}$, $S_{\rm{vN}}^{(1)}$ saturates at finite values, while there is 
a maximum slightly to the right of the phase boundary between the MC and the AC+MC phase (indicated by the dashed line).
 
The difference  $S_{\rm{vN}}^{(1)}(\epsilon_{\rm m}\to \infty) - S_{\rm{vN}}^{(1)}(\epsilon_{\rm m}\to -\infty)$ between the  values for $\epsilon_{\rm m} \ll -t_{\rm a}$  and $\epsilon_{\rm m}\gg t_{\rm a}$ 
 can be estimated from calculating the von Neumann entropy of a pure MC or a pure AC condensate since very large $\epsilon_{\rm m}$ 
fully suppresses the molecules in the AC+MC phase.
In the non-interacting case and for  a bipartition of the system into blocks A and B with lengths $L_{\rm A} = 1$ and $L_{\rm B} = L-1$, $S^{(1)}_{\rm vN}$ can be calculated analytically, yielding for $N$ particles \cite{ding09}
\begin{gather}
S_{\rm vN}^{(1)} = - \sum\limits_{\alpha=0}^N w_{\alpha} \ln{w_{\alpha}} \label{eq:svn_analytic} \\
w_{\alpha} = L^{-N} \binom{N}{\alpha} (L-1)^{(N-\alpha)}. \label{eq:weight_analytic}
\end{gather}
This results in
\begin{eqnarray}
\lim_{\epsilon_{\rm m} \to -\infty} S_{\rm vN}(\epsilon_{\rm m}) &\approx& 1.27  \\
\lim_{\epsilon_{\rm m} \to \infty} S_{\rm vN}(\epsilon_{\rm m}) &\approx& 1.66, \label{eq:svnlimits} 
\end{eqnarray}
where the difference arises from the different filling factors in these limits (namely the corresponding particle numbers are $N=N_{\rm m}=L$ and $N=N_{\rm a}=2L$, respectively).
Those two limiting values overestimate the actual numerical values due to the non-zero repulsive interactions $U_{\rm a}/2=U_{\rm m}/2=U>0$ that lead
to a condensate depletion.
We compare the data in Fig.~\ref{fig:nm_svn_gs_mc_acmc}(b) to the numerical value for a Bose-Hubbard model on a lattice with $L=64$ sites and the corresponding interaction strength: $N=128,~t=1,~U/2=t/3$ for the AC+MC phase and $N=64,~t=0.5,~U/2=t/3$ for the MC phase (plotted in Fig.~\ref{fig:nm_svn_gs_mc_acmc}(b) as dashed lines).
The results for the BHM agree very well with the data for the BBRM in the appropriate limits.

The most striking feature is the system-size dependence of $S_{\rm vN}^{(1)}(\epsilon_{\rm m})$ close to the transition at $\epsilon_{\rm m}^c$ (vertical dashed line in Fig.~\ref{fig:nm_svn_gs_mc_acmc}).
This translates into a pronounced maximum in the derivative of $S_{\rm vN}^{(1)}(\epsilon_{\rm m})$ with respect to $\epsilon_{\rm m}$
in the vicinity of  $\epsilon_{\rm m}^c$, which is plotted in Fig.~\ref{fig:nm_svn_gs_mc_acmc}(c).
With increasing system size $L$ the maximum value grows and an extrapolation to large system sizes via $1/L \to 0$ leads to a finite value for the maximal value in an infinite system.
This suggests that $S_{\rm vN}^{(1)}(\epsilon_{\rm m})$ is sensitive to this phase transition.

The behavior of $S_{\rm vN}^{(1)}$ and of its derivative in the vicinity of $\epsilon_{\rm m}^{\rm c}$ can be understood in the limit of  $U/t_a \rightarrow 0$.
In this case (note that $g=U$), the system is described by the Hamiltonian
\begin{align}
H = &-t_a \sum\limits_j (a_j^{\dagger} a_{j+1} + \text{h.c.}) \notag\\
&-t_m \sum\limits_j (m_j^{\dagger} m_{j+1} + \text{h.c.}) \\
&+\epsilon_m \sum\limits_j n_j^m. \notag
\end{align}
Thus, the ground state is either a condensate of atoms or of molecules depending on the value of $\epsilon_m$.
Since the two species cannot  mix for $g=0$, the transition happens abruptly at a critical detuning $\tilde \epsilon_{m} = -3 t_a$ for $t_m = t_a/2$,
where the system goes from $N_T=N_a=2L$ to $N_T=2N_m=2L$.
This leads to a sudden jump in the local von Neumann entropy and therefore, a singularity  in its derivative with respect to the detuning at $\tilde \epsilon_m$.
The effect of finite interactions is to smoothen this jump, which leads to a finite value for the maximum of the derivative and also a
 shift of the critical point to a smaller value than $\tilde \epsilon_{m} = -3 t_a$.

\subsubsection{MI to MC phase}

\begin{figure}[t]
\includegraphics[width=.96\columnwidth]{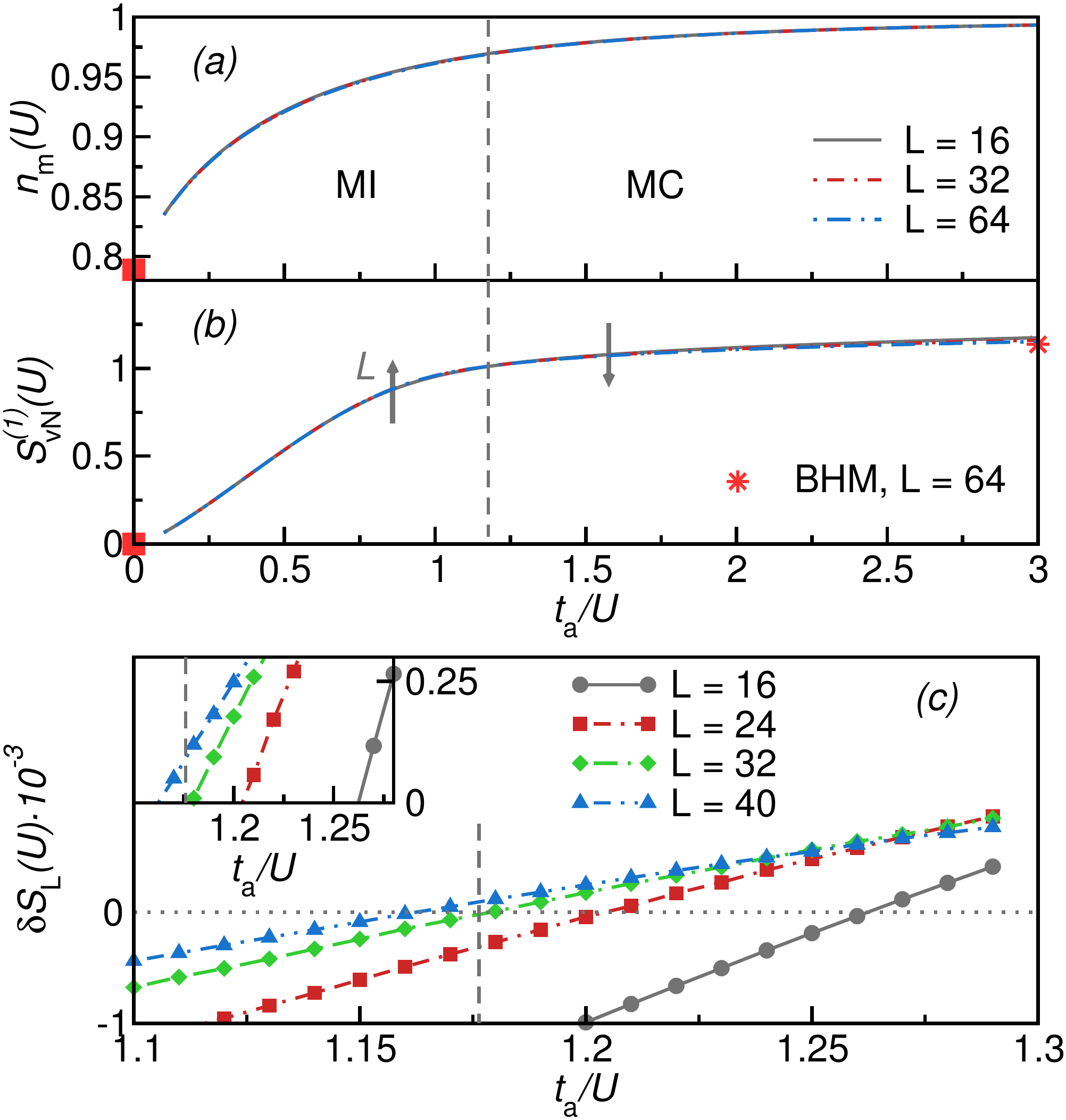}
\caption{(Color online)
(a) Density of molecules and (b) single-site von Neumann entropy in the ground state along the trajectory between the MC and MI phases (see the vertical line in Fig.~\ref{fig:phase-diag} connecting $(\epsilon_{\rm m}= -6t_{\rm a}, \frac{t_{\rm a}}{U} = 3)$ to ($\epsilon_{\rm m}= -6t_{\rm a}, \frac{t_{\rm a}}{U} = 0.1$)).
The arrows in (b) indicate the change of this quantity with increasing system size.
(c) Difference of local von Neumann entropy between systems with $2L$ and $L$ (Eq.~\eqref{eq:svndiff}) as a function of inverse interaction $t_{\rm a}/U$.
The dashed line indicates the phase boundary \cite{ejima11}. The squares in (a) and (b) are the values  for a fully local Hamiltonian Eq.~\eqref{hamil_local}
while the star in (b) is for a Bose-Hubbard model ($L=64$) and appropriately chosen parameters.
The data are calculated using DMRG with  a local cutoff $N_{\rm T}^{(1)} = 30$ and a bond dimension $D=400$.
}
\label{fig:nm_svn_gs_mc_mi}
\end{figure}

Figure~\ref{fig:nm_svn_gs_mc_mi} shows the density of molecules and the von Neumann entropy for the second trajectory which connects the MC and MI phases (compare Fig.~\ref{fig:phase-diag}).
Along this line, $\epsilon_{\rm m}= \text{const}$ while $t_a/U$ is varied.
The density of molecules $n_{\rm m}$ is a monotonically increasing function of $t_a/U$ [see Fig.~\ref{fig:nm_svn_gs_mc_mi}(a)].
The value of $n_{\rm m}$,  in the limiting case of $t_a/U \gg 1 $, is $n_{\rm m} =1$, the maximum possible one for the chosen filling of $N_{\rm T}/L=2$.

The molecular density  and its dependence on $\epsilon_{\rm m}$ can also be understood in the limit of weak interactions $t_{\rm a}/U \to \infty$ 
(in this limit, $g=0$ as well). As discussed above there is a critical $\tilde \epsilon_{\rm m}$ for which one obtains 
either molecules ($\epsilon_{\rm m} <\tilde \epsilon_{\rm m}$) or atoms ($\epsilon_{\rm m} >\tilde \epsilon_{\rm m}$) only, with $\tilde \epsilon_{\rm m}=-3t_{\rm a}$. 
For the parameters of Fig.~\ref{fig:nm_svn_gs_mc_mi}, we have $\epsilon_{\rm m}=-6t_{\rm a}$ and hence mostly molecules for $t_{\rm a}/U \gg 1$.

In the limit of large $U/t_{\rm a}$, the Hamiltonian is fully local and {\it blockdiagonal} in the total local particle number $N_{\rm T}^{(1)}$,
\begin{eqnarray}
\text{H} &=& U \sum\limits_{j} [ n_{j,\rm a}(n_{j,\rm a} - 1) + n_{j,\rm m}(n_{j,\rm m} - 1) \label{hamil_local} \\
&&+ n_{j,a}n_{j,m} + m_{j}^{\dagger} a_{j} a_{j} + m_{j} a_{j}^{\dagger} a_{j}^{\dagger} ]. \notag
\end{eqnarray}
The ground state in the $N_{\rm T}^{(1)}=2$ subspace and for $g=0$ is thus a product state, with the local state consisting of one molecule per site:
\begin{eqnarray}
|\psi^{\rm{MI},g=0}_0\rangle &=& \prod_{j} |\phi_0^{\rm{MI},g=0}\rangle_j \\
\quad |\phi_0^{\rm{MI},g=0}\rangle_j &=& |0;1\rangle.
\end{eqnarray}
A nonzero $g$ can only couple the states in the $N_{\rm T}^{(1)}=2$ sector and thus mixes in the state with two atoms.
For the case of $g=U$ (and setting $\epsilon_{\rm m}=0$), we obtain:
\begin{equation}
 |\phi_0^{\rm{MI},g=U}\rangle = \frac{1}{\sqrt{3-\sqrt{3}}} \left( \frac{1-\sqrt{3}}{\sqrt{2}} \ket{2;0} + \ket{0;1} \right). \label{eq:mi_gs}
\end{equation}
Using this state, we 
can calculate the molecular density in this limit, finding $n_{\rm m}(U/t_{\rm a} \gg 1)\approx 0.789$, which is in perfect agreement with the
numerical results shown in Fig.~\ref{fig:nm_svn_gs_mc_mi}(a) when extrapolated to $t_a/U =0$.

Figure~\ref{fig:nm_svn_gs_mc_mi}(b) shows the single-site von Neumann entropy as a function of $t_{\rm a}/U$.
In the limit of $U/t_{\rm a}=0$, the system is in a (non-interacting) molecular condensate.
Hence, the von Neumann entropy approaches the value $S_{\rm{vN}}^{(1)} = 1.27$ [see Eq.~(\ref{eq:svnlimits})] with decreasing $U/t_{\rm a}$ [indicated by the horizontal dotted line in Fig.~\ref{fig:nm_svn_gs_mc_mi}(b)].
In the opposite limit, the Hamiltonian is fully local and the entanglement entropy thus vanishes.
In between these limits, $S_{\rm vN}^{(1)}$ is a monotonically increasing function of $t_{\rm a}/U$.

Contrary to Fig.~\ref{fig:nm_svn_gs_mc_acmc}(b) there is no directly obvious feature in $S_{\rm vN}^{(1)}$ at the phase transition.
$S^{(1)}_{\rm vN}$ exhibits a maximum in its first derivative which, however, converges to a point far below $(U/t_{\rm a})_c=1.176$ as $L$ increases and this maximum is therefore not connected to the phase transition between the superfluid and the MI phase.
In fact, previous studies \cite{giorda04,buonsante07} of the Bose-Hubbard model at unit filling found a similar behavior of the single-site entanglement entropy.

However, comparing the curves for different system sizes, we find that  $S_{\rm vN}^{(1)}$  is a monotonically increasing(decreasing) function of $L$ in the MI(MC) phase (see the arrows in Fig.~\ref{fig:nm_svn_gs_mc_mi}(b)).
The qualitative behavior in the MI phase can be explained as a consequence of the finite (exponentially decaying) correlation length \cite{ejima12}.
Because of that, the local entropy has to increase with system size until it saturates at a finite value when the system is large enough to support the full correlation length.
This behavior can be verified numerically for a point deep in the MI phase for both BBRM and BHM.

The behavior in the MC phase can be obtained by calculating the first derivative of the local von Neumann entropy Eq.~\eqref{eq:svn_analytic} with respect to system size $L$ explicitly in leading order in $1/L$  in a non-interacting condensate.
In order to be able to do this analytically we use an upper bound for the binomial coefficient $\binom{n}{k} \le n^k/k!$ (for details, see the Appendix~\ref{app:perttheory}).
We find, to leading order, $S_{\rm vN}^{(1)} =a+ b/L$, where $a,b$ are constants.
Again, choosing a point deep in the MC phase this behavior can be verified both for the BBRM and BHM.

This suggests to study the $L$-dependence of the difference between two curves for system sizes $L$ and $2L$
\begin{align}
\delta S_{L}(U) = S_{\rm{vN},L}^{(1)}(U) - S_{\rm{vN},2L}^{(1)}(U). \label{eq:svndiff}
\end{align}
This choice ensures that the ratio between system sizes is kept constant.
Interestingly, the difference in local entanglement entropy is a linear function of $t_{\rm a}/U$ in the vicinity of the phase transition (see Fig.~\ref{fig:nm_svn_gs_mc_acmc}(c)).
The point where the curves cross zero are the points where the monotony of the local entropy changes as a function of $L$.
This corresponds to the observation that the local entropy increases in the MI phase, while it decreases in the MC phase.
With increasing system size two effects occur.
The slope decreases and the point where the curve hits zero shifts to the left, in the direction of the phase boundary (indicated by the dashed line in Fig.~\ref{fig:nm_svn_gs_mc_acmc}(c) \cite{ejima11}).
A naive extrapolation to $1/L = 0$, however, yields  an estimate for the critical point that is below the literature value.
We also study the BHM at unit filling which corresponds to the $\epsilon_{\rm m} \rightarrow -\infty$ limit of the BBRM for our choice of parameters.
In this case, the local von Neumann entropy (and  $\delta S_{L}(U)$ as defined in Eq.~\eqref{eq:svndiff}) 
behaves in exactly the same way as in the BBRM, as expected.

\subsection{Structure of optimal modes}
\label{sec:gs_w_om}

In this section we study the structure of optimal modes (defined in Sec.~\ref{sec:optmodes}) for specific points in each one of the three phases and then we investigate the changes in these states as the phase boundaries are crossed.

\subsubsection{MI phase}

\begin{figure}[t]
\includegraphics[width=.96\columnwidth]{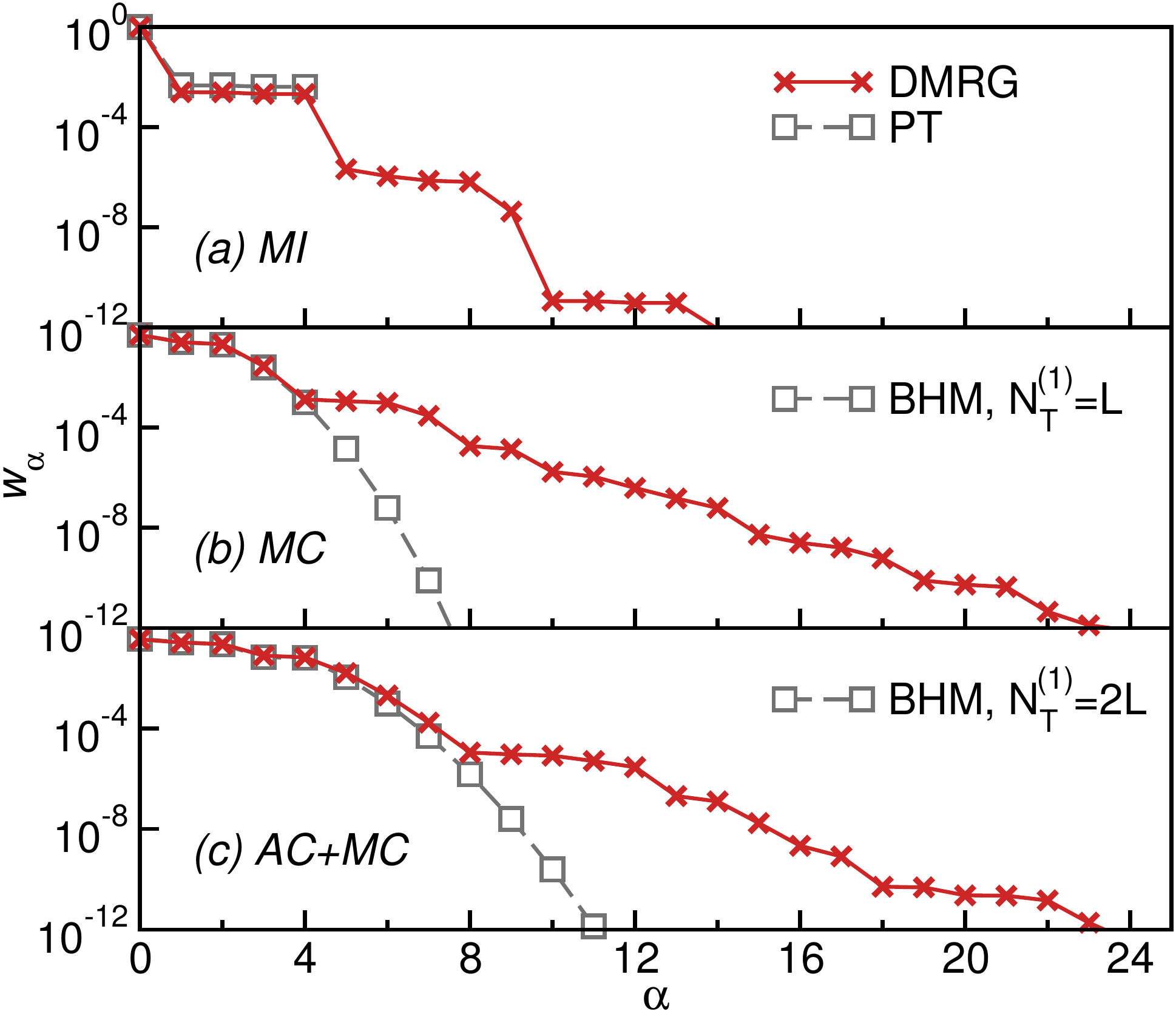}
\caption{(Color online)
Weights of the optimal modes in (a) MI ($\epsilon_{\rm m} = -6t_{\rm a}, \frac{t_{\rm a}}{U} = 0.1$), (b) MC ($\epsilon_{\rm m} = -6t_{\rm a}, \frac{t_{\rm a}}{U} = 3$)
and (c) AC+MC ($\epsilon_{\rm m} = 2t_{\rm a}, \frac{t_{\rm a}}{U} = 3$) phase.
Closed symbols: DMRG for $L=64$.
Open symbols: in (a) perturbative results, (b) Bose-Hubbard model with $N=L=64$ and (c) $N=2L=128$.
}
\label{fig:w_gs}
\end{figure}

We start with the MI phase, considering the parameters   $\frac{t_{\rm a}}{U} = 0.1, \epsilon_{\rm m} = -6t_{\rm a}$.
We first take a look at the weight spectrum $w_\alpha$ shown in Fig.~\ref{fig:w_gs}(a).
The spectrum is dominated by the first optimal mode with $w_0 \lesssim 1$.
Then there are sequences of plateaus of several states each with very similar weights.
In the first of these plateaus, there are two pairs of states that are very close to each other.

\begin{figure}[t]
\includegraphics[width=.96\columnwidth]{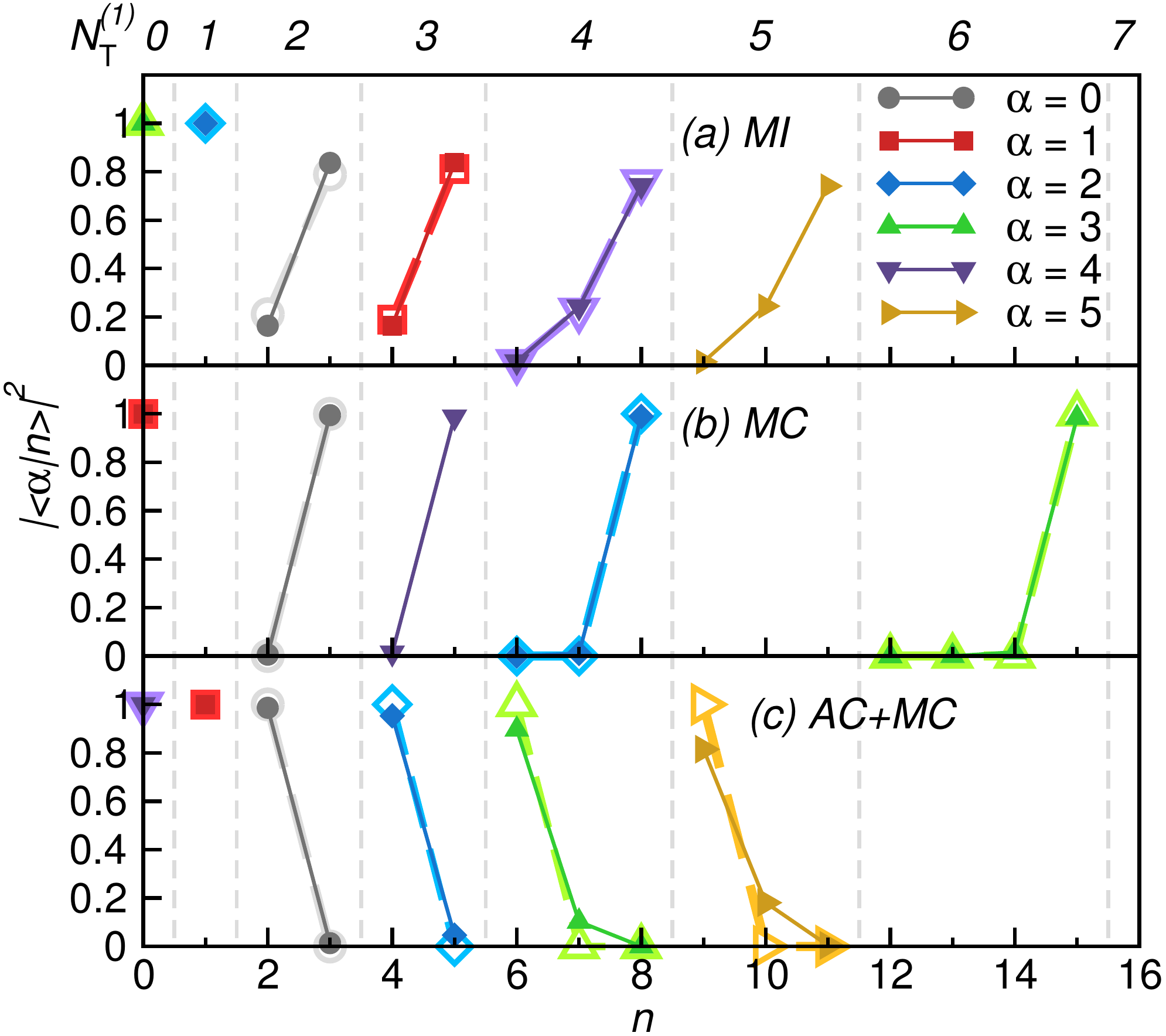}
\caption{(Color online)
The optimal modes in (a) MI ($\epsilon_{\rm m} = -6t_{\rm a}, \frac{t_{\rm a}}{U} = 0.1$), (b) MC ($\epsilon_{\rm m} = -6t_{\rm a}, \frac{t_{\rm a}}{U} = 3$) and (c) AC+MC ($\epsilon_{\rm m} = 2t_{\rm a}, \frac{t_{\rm a}}{U} = 3$) phase. Filled symbols: numerical data (DMRG, $L=64$), open symbols: perturbative results (MI) or estimate from a BHM with $N=L$ (MC) or $N=2L$ (AC+MC).
The label $n$ is defined in Tab.~\ref{tab:localbasis}.
The vertical dashed lines along with the label $N_{\rm T}^{(1)}$ indicate in which particle number subsector the states with label $n$ are located.
}
\label{fig:m_gs}
\end{figure}

The optimal-mode spectra are shown in Fig.~\ref{fig:m_gs}(a).
First, we see that the optimal modes are very simple superpositions of the bare modes.
They are constrained in form because, as discussed above, they can only mix bare states with the same total particle number $N_{\rm T}^{(1)}$.
Therefore, the more interesting modes are the ones that mix bare modes, e.g., $\alpha=0,1,4,5$ where the relative contributions of the bare modes in the large $U/t_{\rm a}$ limit depend on the parameters $\epsilon_{\rm m}$ and $g$ (the latter being tied to $U$ in our study).

More information can be obtained from perturbation theory in the hopping parameter $t_{\rm a}$ (remember that in our case, $t_{\rm m} = t_{\rm a}/2$).
From the previous discussion (Sec.~\ref{sec:gs_nm_svn}), we know that in the $t_{\rm a}/U=0$ limit the ground state is a product state of the local state given in Eq.~(\ref{eq:mi_gs}).
Therefore, for exactly $t_{\rm a}/U=0$
\begin{equation}
\ket{\alpha = 0} = \ket{\phi_0^{\rm{MI},g=U}}. \label{eq:optmode0}
\end{equation}
A first approximation to the weight spectrum of the first five modes and the structure of the four next-to-leading modes ($\alpha=1,2,3,4$) for finite but small $t_{\rm a}, t_{\rm m} \ll U$ can be calculated when considering the first-order correction terms of the wave function
\begin{align}
| \tilde{\psi}_0  \rangle & \propto | \psi_0^{(0)} \rangle + \frac{t_{\rm a}}{U} | \psi_0^{(1)} \rangle \notag \\
\rho^{(1)} &\approx {\rm tr_E}(| \tilde{\psi}_0 \rangle \langle \tilde{\psi}_0 |).
\end{align}

The weight spectrum and structure calculated from perturbation theory (PT) are shown in Figs.~\ref{fig:w_gs}(a) and \ref{fig:m_gs}(a) as open symbols.
Both are very close to the numerically exact data because we are very deep in the large $U/t_{\rm a}$ limit.
The plateau structure in Fig.~\ref{fig:w_gs}(a) emerges in first-order perturbation theory.
Qualitatively, the structure of the weight spectrum and the optimal-mode spectra can be understood from the following observation.
In first order the atomic hopping term couples the optimal mode in the $t_{\rm a}/U = 0$ limit (see Eq.~\eqref{eq:mi_gs}) which resides in the $N_{\rm T}^{(1)} = 2$ subspace to a state in the $N_{\rm T}^{(1)} = 1$ subspace and a state in the $N_{\rm T}^{(1)} = 3$ subspace.
By contrast, the molecular hopping term couples this state to a state in the $N_{\rm T}^{(1)} = 0$ subspace and a state in the $N_{\rm T}^{(1)} = 4$ subspace.
Because of that there are two pairs of degenerate optimal modes.
This perfectly describes the relative positions of the optimal-mode spectra of the modes $\alpha=1,2,3 \text{ and } 4$ in Fig.~\ref{fig:m_gs}(a).
Following this reasoning we can also qualitatively explain the second and third plateaus in the spectrum.
In second order the system couples to the states in the $N_{\rm T}^{(1)} = 0,1,2,3,4,5,6$ subspaces.
Since the $N_{\rm T}^{(1)} = 0,1$ subspaces are non-degenerate and contain only one state, they do not contribute to the second plateau.
By comparison to the numerics we find that the second plateau corresponds to exactly the remaining five states originating from the $N_{\rm T}^{(1)} = 2,3,4,5,6$ subspaces.
The same argument holds also for the third  plateau present in Fig.~\ref{fig:w_gs}(a).

\subsubsection{MC and AC+MC phase}

\begin{figure}[t]
\includegraphics[width=.96\columnwidth]{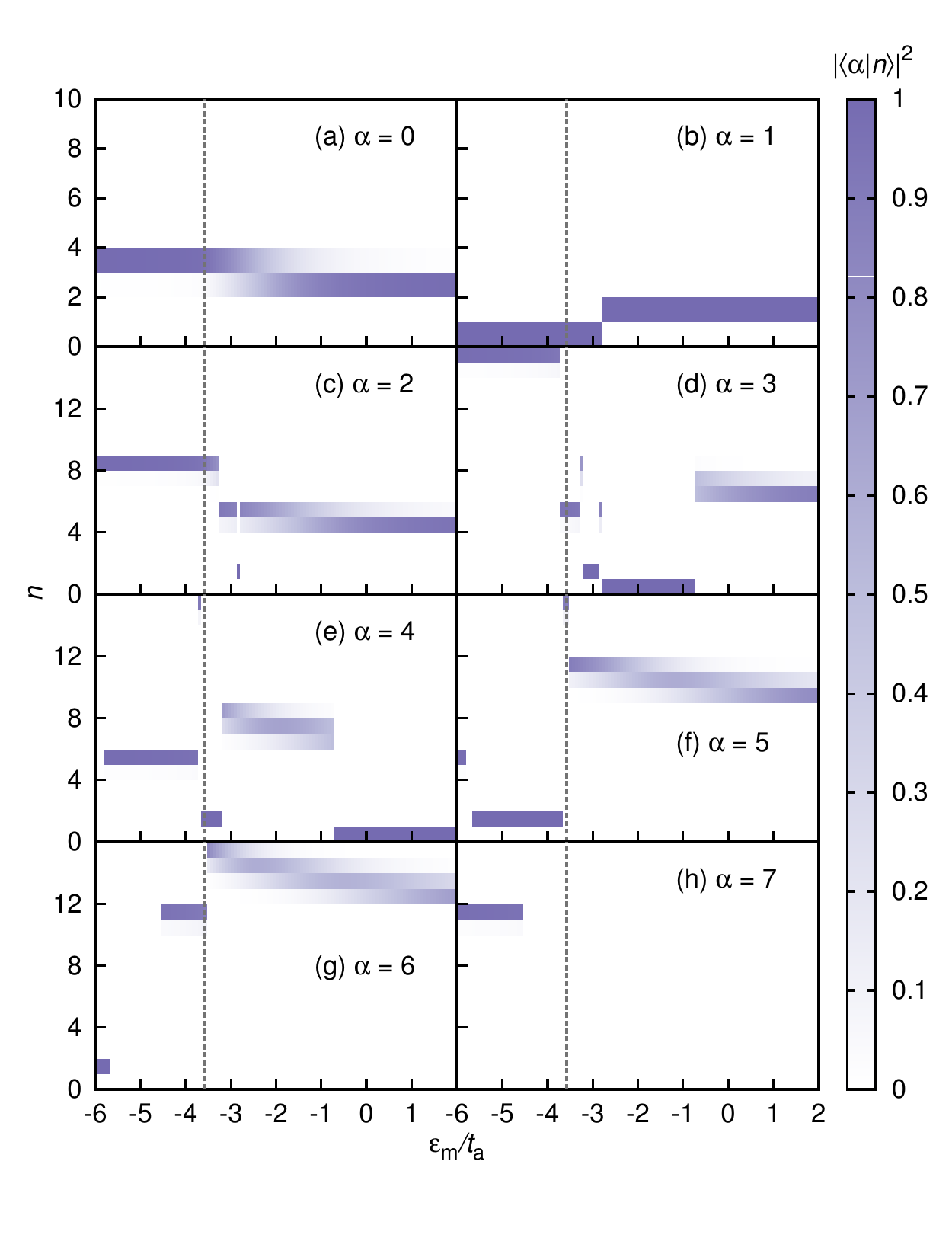}
\caption{(Color online)
(a)-(h) Evolution of the first eight optimal modes $|\alpha \rangle $ along the trajectory from the  MC ($\epsilon_{\rm m} = -6t_{\rm a}, \frac{t_{\rm a}}{U} = 3$) to the AC+MC phase ($\epsilon_{\rm m} = 2t_{\rm a}, \frac{t_{\rm a}}{U} = 3$).
The figure shows the weights $|\langle \alpha | n\rangle|^2$ of the bare local states $|n\rangle=|N_{\rm a}^{(1)}(n),N_{\rm m}^{(1)}(n)\rangle$
contributing to the optimal modes as a function of detuning. DMRG results for $L=80$.
The dashed line indicates the phase boundary \cite{ejima11}.
The respective physical state corresponding to the index $n$ is defined in Tab.~\ref{tab:localbasis}.
}
\label{fig:modes2}
\end{figure}

Next, we consider the weight spectrum and optimal modes  for the MC and AC+MC phases which are illustrated in Figs.~\ref{fig:w_gs}(b),(c) and \ref{fig:m_gs}(b),(c) respectively.
The weights for the MC and AC+MC phases [Figs.~\ref{fig:w_gs}(b) and (c)] are compared to those computed for a Bose-Hubbard model at unit (MC) and double filling (AC+MC).
In the MC phase we compare to a BHM with $N_{\rm T} = L$ particles because for our choice of parameters,  the number of atoms is negligible (see Fig.~\ref{fig:nm_svn_gs_mc_acmc}(a)).
In the AC+MC phase we compare to a BHM with $N_{\rm T} = 2L$ particles since in this phase very few molecules are present (again consult Fig.~\ref{fig:nm_svn_gs_mc_acmc}(a)) and thus, all the particles are unbound.
The weights are computed using DMRG for a system of size $L=64$.

These estimates provide a very good approximation of the exact values until they begin to deviate at $w_{\alpha} \approx 10^{-4}$ where the weights start to decay slower in the BBRM compared to the 
(single-component)  BHM.
The reason is that there are more than just one species present in both states, plus effects of the Feshbach term.

Figure~\ref{fig:m_gs}(b) shows the optimal-mode spectra for a point in the  MC phase.
The individual optimal modes in this phase are the bare occupation number states, with virtually no mixing in the degenerate subspaces (e.g., $N_{\rm T}=2$).
The first noticeable deviation between the prediction for the optimal-mode structure from considering a BHM compared to the 
numerical data occurs for the $\alpha = 4$ optimal mode: this mode consists of a molecule and an atom,
showing that the presence of atoms is still important for this state.
The simple structure of the optimal modes in the MC phase roots in the fact that adding an atom is suppressed for this choice of $\epsilon_{\rm m}$.

In the AC+MC phase we observe a very similar behavior: the first two modes are perfectly peaked.
The other ones are already mixtures of atoms and molecules which shows that one can not neglect one of those species in this phase.

\begin{figure}[t]
\includegraphics[width=.96\columnwidth]{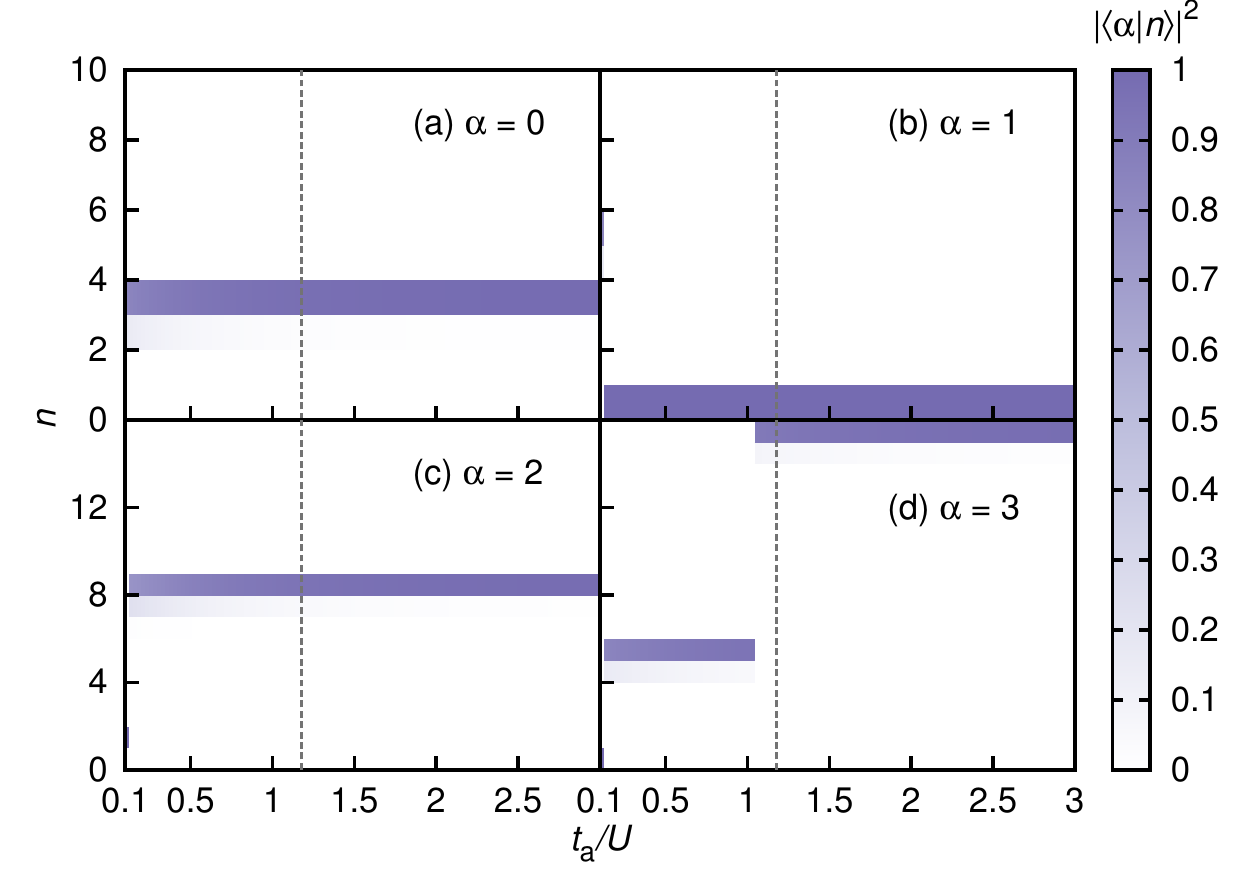}
\caption{(Color online)
(a)-(d) Evolution of the first four  optimal modes $|\alpha \rangle $ along the trajectory from the MI ($\epsilon_{\rm m} = -6t_{\rm a}, \frac{t_{\rm a}}{U} = 0.1$) to the MC phase ($\epsilon_{\rm m} = -6t_{\rm a}, \frac{t_{\rm a}}{U} = 3$).
The figure shows the weights $|\langle \alpha | n\rangle|^2$ of the bare local states $|n\rangle=|N_{\rm a}^{(1)}(n),N_{\rm m}^{(1)}(n)\rangle$ contributing to the optimal modes as a function of $t_{\rm a}/U$.
DMRG results for $L=64$.
The dashed line indicates the phase boundary \cite{ejima11}.
The respective physical state corresponding to the index $n$ is defined in Tab.~\ref{tab:localbasis}.
}
\label{fig:modes1}
\end{figure}

\subsection{Evolution of optimal modes across phase transitions}

In this section we discuss the optimal-mode structure when tuning the system continuously crossing one of two phase boundaries, i.e., either the MC to AC+MC transition or the MI to MC transition.

\subsubsection{The MC to AC+MC transition}

As discussed before, the local entanglement entropy is sensitive to the location of this transition (Sec.~\ref{sec:gs_nm_svn}).
It is very curious to ask how the optimal-mode structure changes as a function of detuning $\epsilon_{\rm m}$.
Figures~\ref{fig:modes2}(a) - \ref{fig:modes2}(h) show the projection of the most important eight optimal modes on the $n$-th bare mode as a function of detuning $\epsilon_{\rm m}$.
Generally, all eight states seem to change significantly in the vicinity of the phase boundary.
This is  expected by inspection of  Fig.~\ref{fig:m_gs}: the optimal-mode structures at the two points deep in the phases are different and at some point,
a reorganization has to occur.
Also, we see that this transition manifests itself in one of two ways:

{(i) \it A continuous transition.}
When the two modes are located in the same block (i.e., they can be labeled with the same total number of particles $N_{\rm T}^{(1)}$) the transition is smooth (excluding the case of $U = 0$).
An example for this behavior is given in Fig.~\ref{fig:modes2}(a):
We see in Fig.~\ref{fig:m_gs} that the most important optimal mode in the MC phase is in the $N_{\rm T}^{(1)} = 2$ block where the two atoms are bound inside a molecule.
The corresponding optimal mode in the AC+MC phase lies in the same block but is a state where the two particles are unbound.

{(ii) \it A level crossing.}
When two modes are located in different blocks a sudden jump can occur when the weights cross each other.
An example for this behavior is shown in Fig.~\ref{fig:modes2}(b):
In this case the modes can not smoothly transform into each other and the structures stay roughly the same until their weights suddenly swap position and, therefore, one mode becomes more important than the other one.
Since the weights are a smooth function of the detuning (at least for our system size) this means that a mode that has a low weight in the initial state and a high one in the final state has to climb until it reaches its final position.
Examples for this behavior are shown in Figs.~\ref{fig:modes2}(b),(c) and (d):
The $N_{\rm T}^{(1)}=1$ state has a low weight in the MC phase and thus has to ascend the ladder of optimal states 
until it reaches its final position as the second-most important mode in the AC+MC phase state.
Its ascent can first be seen in Fig.~\ref{fig:modes2}(d) where this state first gains appreciable weight
while  upon further increasing $\epsilon_{\rm m}$, it moves to Fig.~\ref{fig:modes2}(c) where it stays only shortly until it reaches its final position in Fig.~\ref{fig:modes2}(b) for all larger values of the detuning.

So far, we have discussed how two different modes can transform directly into another.
We want to get some insight into what happens to the optimal modes when crossing the phase boundary.
Figure~\ref{fig:modes2}(a) shows how the $\alpha = 0$ mode evolves during the transition: for small detuning the majority of the weight is in the $| n = 3 \rangle = | 0,1 \rangle$ state.
As the detuning increases the weight gets shifted over to the $| n = 2 \rangle = | 2,0 \rangle$ state.
Thereby, without having any other information we can conclude that the system favors molecules in one phase and atoms in the other.
The same happens in all other modes where the optimal modes mix more than one state.
An interesting feature emerges in the higher optimal modes $\alpha = 5,6$ (Figs.~\ref{fig:modes2}(f),(g)): the corresponding optimal modes change their 
structure abruptly by jumping from (linear combinations of) small to large $n$ states.
 The occurrence of this jump  is independent of system size and sits right at the phase boundary.
We note that most rearrangements in the optimal modes, independent of their nature, happen in the vicinity of the phase boundary (indicated by the dashed lines in the figures) and thereby the changes in the optimal modes are correlated to this transition.

\subsubsection{The MI to MC transition}

The second transition that we study is the one from the MI to the MC phase by varying the interaction $U$ only.
The corresponding results are shown in Figs.~\ref{fig:modes1}(a) - \ref{fig:modes1}(d).
By inspecting Fig.~\ref{fig:m_gs}, we see that apart from the first optimal mode no pair of them is located in the same block and thus they all have to undergo a level crossing.
The change in the second and third mode shown in Figs.~\ref{fig:modes1}(b),(c) occurs at a value close to $t_{\rm a}/U = 0.1$.
Again, the $\alpha = 0$ mode lies in the $N_{\rm T}^{(1)} = 2$ subspace for the whole range of observed values of $t_{\rm a}/U$.
With increasing $t_{\rm a}/U$ the atomic contribution to this mode gets suppressed because the detuning term dominates the occupation ratio.
Another feature in the vicinity of the phase transition is shown in Fig.~\ref{fig:modes1}(d): a jump occurs where the $N_{\rm T}^{(1)} = 3$ mode drops to an even lower weight and the $N_{\rm T}^{(1)} = 6$ mode moves to its final relative position.
Again, the optimal modes show features in the vicinity of the phase boundary which suggest that they are sensitive to this transition as well.

\section{Quantum quenches}
\label{sec:ququenches}

In the last section we calculated observables in the ground state along two trajectories in the phase diagram where both crossed a phase boundary (see Fig.~\ref{fig:phase-diag}).
This procedure can be seen as evolving the system in time from one point in the phase diagram to another one adiabatically.
In this section we change parameters instantaneously between the points marked in Fig.~\ref{fig:phase-diag} via a quantum quench.

\begin{figure}[t]
\includegraphics[width=.96\columnwidth]{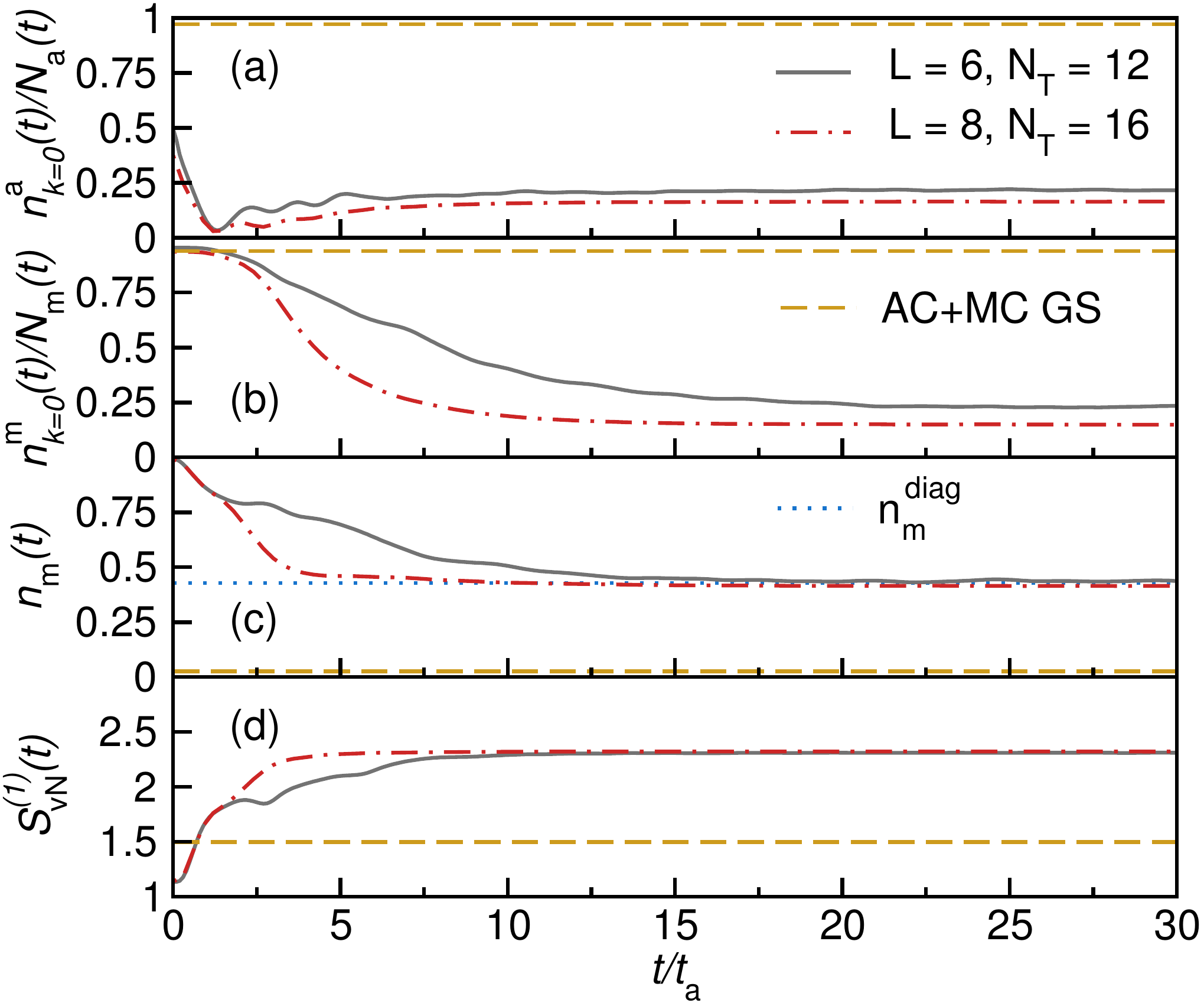}
\caption{(Color online)
Evolution of the $k=0$ component of the (a) atomic and (b) molecular momentum distribution function, (c) the  molecular density and (d) the single-site von Neumann entropy as a function of time along the trajectory from the MC ($\epsilon_{\rm m} = -6t_{\rm a}, \frac{t_{\rm a}}{U} = 3$) to the AC+MC phase ($\epsilon_{\rm m} = 2t_{\rm a}, \frac{t_{\rm a}}{U} = 3$).
The dashed line in (c) gives the expectation value of the molecular density in the diagonal ensemble (Eq.~\eqref{eq:diagens}) for a system of size $L=6$ which shows that we reach the long-time steady state in the observed time.
The quench energy is $E_{\rm q}/L = (E-E_0)/L = 4.1227$.
ED results for $L=6,8$.
}
\label{fig:nk_nm_svn_t_1}
\end{figure}

\begin{figure}[t]
\includegraphics[width=.96\columnwidth]{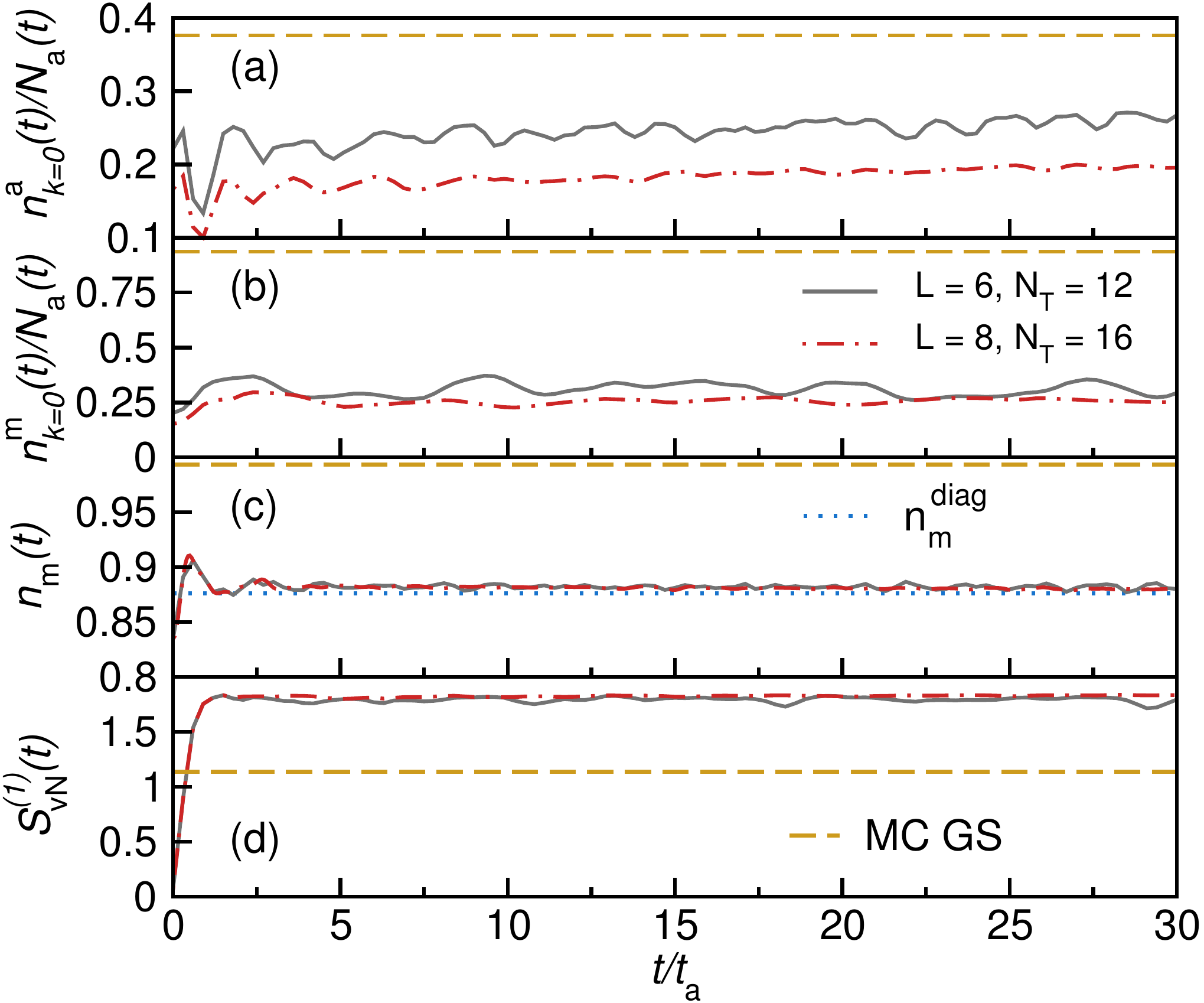}
\caption{(Color online)
 Evolution of the $k=0$ component of the (a) atomic and (b) molecular momentum distribution function, (c) the  molecular density and (d) the single-site von Neumann entropy as a function of time along the trajectory from the MI ($\epsilon_{\rm m} = -6t_{\rm a}, \frac{t_{\rm a}}{U} = 0.1$) to the MC phase ($\epsilon_{\rm m} = -6t_{\rm a}, \frac{t_{\rm a}}{U} = 3$).
The dashed line in (c) gives the expectation value of the molecular density in the diagonal ensemble (Eq.~\eqref{eq:diagens}) for a system of size $L=6$ which shows that we reach the long-time steady state in the observed time.
The quench energy is $E_{\rm q}/L = (E-E_0)/L = 1.3923$.
ED results for $L=6,8$.
}
\label{fig:nk_nm_svn_t_2}
\end{figure}

\subsection{Number of molecules and single-site von Neumann entropy}
\label{sec:quench_nmsvn}

For the quenches, in addition to the real-space observables $n_{\rm m}$ and $S_{\rm{vN}}^{(1)}$,
we also study the $k=0$ component of the quasimomentum distribution function of atoms (molecules) rescaled by the total number of atoms (molecules).
Also, we calculate the long-time limit of the expectation value of the molecular density which is given by its expectation value in the so-called diagonal ensemble \cite{rigol08}
\begin{align}
\overline{\langle n_{\rm m} \rangle}_{\rm diag} = \sum\limits_{n} |\langle \psi_0 | \psi_n \rangle|^2 \langle \psi_n |n_{\rm m}| \psi_n \rangle, \label{eq:diagens}
\end{align}
where $|\psi_n \rangle$ are the eigenstates of the postquench Hamiltonian and $| \psi_0 \rangle$ is the initial state before quenching.
Additionally, we compare the expectation value of the molecular particle number density in the diagonal ensemble with the one in the canonical ensemble and of the local von Neumann entropy in the steady state with the one in the canonical ensemble. For the calculation of expectation values in the diagonal and canonical ensembles 
we use a system of size $L=6$ due to the need of a full diagonalization of the Hamiltonians.

\subsubsection{Quench between the MC and the AC+MC phase}

The dynamics in the first quench from the MC to the AC+MC phase is illustrated in Fig.~\ref{fig:nk_nm_svn_t_1} for system sizes $L=6,8$.
In the first few time steps the $k=0$ quasimomentum occupations of both species - atoms and molecules - decrease.
The decrease of $n_{\rm m}^{k = 0}$ is consistent with the behavior of the molecular density $n_{\rm m} = N_{\rm m}/L$: it decreases in time which means that atoms are created.

We define the quench energy as
\begin{equation}
E_q = E-E_0; \quad E= \langle \psi_0 | H | \psi_0\rangle
\end{equation}
where $H$ is the postquench Hamiltonian. The quench energy in this quench 
is finite and so large  that the initial state samples primarily eigenstates in the bulk of the spectrum and it is therefore not surprising that the observables are not comparable to their ground-state expectation values.

We also calculate the diagonal and canonical ensemble average for the molecular density in a system of size $L=6$ (diagonal ensemble: dotted line in Fig.~\ref{fig:nk_nm_svn_t_1}(c)) \cite{sorg14}.
For the calculation of the canonical expectation value we first extract the canonical temperature $T$ by fixing the expectation value of the energy in the canonical ensemble to the energy of the initial state with respect to the postquench Hamiltonian
\begin{align}
\langle H \rangle_{\rm can}(\beta) = \frac{\sum\limits_n E_n \text{e}^{-\beta E_n}}{\sum\limits_n \text{e}^{-\beta E_n}} \overset{!}{=} \langle \psi(t=0) | H | \psi(t=0) \rangle.
\end{align}
Expectation values of observables $\hat O$ in the canonical ensemble are computed from:
\begin{equation}
\langle \hat O \rangle= \mbox{tr}(\rho_{\rm can} \hat O); \quad \rho_{\rm can} = e^{-\beta H}/Z
\end{equation}
where $Z$ is the partition function and $\beta =1/T$.

The real-time data for both system sizes lie on top of the diagonal ensemble average value which shows that the molecular density has fully relaxed to its
infinite-time value.
The canonical ensemble average is close to the diagonal ensemble average with a relative difference of $(n_{\rm m}^{\rm diag}-n_{\rm m}^{\rm can})/n_{\rm m}^{\rm diag} \approx 9\%$ for $L=8$.
The remaining difference can be attributed to finite-size effects \cite{sorg14}.
As expected, the local von Neumann entropy increases in time up to a point where the system reaches a steady state.
Increasing system size has two effects: first, the atomic zero-quasimomentum occupation  decreases and, second, oscillations in time vanish.
Apart from this, the data for all considered system sizes  agree very well.
We calculate the local von Neumann entropy in the canonical ensemble and find that it deviates from the steady-state value $(S_{\rm vN}^{(1)})_{\rm st}$ by $((S_{\rm vN}^{(1)})_{\rm st}-(S_{\rm vN}^{(1)})_{\rm can})/(S_{\rm vN}^{(1)})_{\rm st} \approx 0.01\%$.

\subsubsection{Quench from the MI to the MC phase}

Figure~\ref{fig:nk_nm_svn_t_2} shows our results for a quench from the MI to the MC phase.
Here, the quench energy is much smaller, probing the postquench spectrum at its lower edge. 
Similar to the previously discussed quench, the changes in all observed quantities occur very rapidly: after a very short transient time a relaxation to a steady-state value occurs.
The zero-quasimomentum occupations of neither the atoms nor the  molecules  change significantly as a function of time.
Also, the molecular density increases only slightly.
The most dramatic change happens in the von Neumann entropy which shows a very steep increase and then stays constant. 
Again, the finite size of the system introduces fast oscillations whose amplitudes decrease  as system size is increased.
The molecular density reaches its long-time limit during the observed time as indicated by the diagonal ensemble average (dashed line in Fig.~\ref{fig:nk_nm_svn_t_2}(c)).
For this quench the relative difference between diagonal and canonical ensembles is $(n_{\rm m}^{\rm diag}-n_{\rm m}^{\rm can})/n_{\rm m}^{\rm diag} \approx 2\%$.
For the local von Neumann entropy, we   find a relative deviation the steady-state value from the canonical ensemble  $((S_{\rm vN}^{(1)})_{\rm st}-(S_{\rm vN}^{(1)})_{\rm can})/(S_{\rm vN}^{(1)})_{\rm st} \approx 5\%$ ($L=8$).

\subsubsection{Post-quench eigenstate expectation values}
\label{sec:spectrum}

The relaxation of the system to a steady state  can be understood from the distribution of diagonal postquench eigenstate expectation values (DPQEV) $O_{nn} = \langle \psi_n|\hat O|\psi_n \rangle$ and the overlaps of the initial state with the eigenstates of the postquench Hamiltonian.
This relates the notion of thermalization in a closed quantum system to the eigenstate thermalization hypothesis (ETH) \cite{rigol08,srednicki94,deutsch91},
which is a widely used concept in this field (see, e.g., \cite{sorg14} and references therein). We summarize its  essence  here.

The overlaps $c_n$ of the initial state with the postquench eigenstates, given by 
\begin{equation}
|\psi(t=0) \rangle = \sum_n c_n |\psi_n\rangle\,,
\end{equation}
determine which eigenstates contribute significantly to the time evolution.
The time evolution of any observable is given by
\begin{align}
\langle \hat{O} \rangle(t) =& \sum\limits_{n} |c_n|^2 \langle \psi_n|\hat{O}| \psi_n \rangle \notag \\
&+ \sum\limits_{nn',n \ne n'} c_n^* c_{n'} \langle \psi_n|\hat{O}|\psi_{n'} \rangle \text{e}^{i(E_n-E_{n'})t}.
\end{align}
For long-time averages the oscillating terms cancel (see the discussion in \cite{rigol08}) and we are left with the time-independent part only 
\begin{align}
\overline{\langle \hat{O} \rangle} = \sum\limits_{n} |c_n|^2 \langle \psi_n|\hat{O}|\psi_n \rangle. \label{eq:longtexpval}
\end{align}
This is the diagonal ensemble.
Up to this point the statements are exact and the above equation has to hold for every observable.
One can now ask when the above defined expectation value coincides with the expectation value computed in   a thermal ensemble.
Since in  our closed quantum system,  energy, particle number and volume are fixed,  it is natural to compare Eq.~\eqref{eq:longtexpval} to the microcanonical expectation value
\begin{align}
\langle \hat{O} \rangle_{\rm m} =& \frac{1}{N} \sum\limits_{E-\Delta E < E_n < E+\Delta E} \langle \psi_n|\hat{O}|\psi_n \rangle \\
\sum\limits_n |c_n|^2 O_{nn} \overset{!}{=}& \frac{1}{\mathcal{N}} \sum\limits_{E-\Delta E < E_n < E+\Delta E} \langle \psi_n|\hat{O}|\psi_n \rangle
\label{eq:eth}
\end{align}
where $\Delta E$ is a small width around the mean energy $E$ and $\mathcal{N}$ gives the number of states with an energy inside that energy window.
The eigenstate thermalization hypothesis (ETH) \cite{deutsch91,srednicki94,rigol08} makes a statement of how the above equation Eq.~\eqref{eq:eth} can be fulfilled:
It will in general work out when (i) the $|c_n|^2$  sample just a very narrow energy region (comparable to $\Delta E$) and (ii) the $O_{nn}$ are a sharp distribution
and thus only a  function of  energy in the region that the $|c_n|^2$ sample.

\begin{figure}[t]
\includegraphics[width=.96\columnwidth]{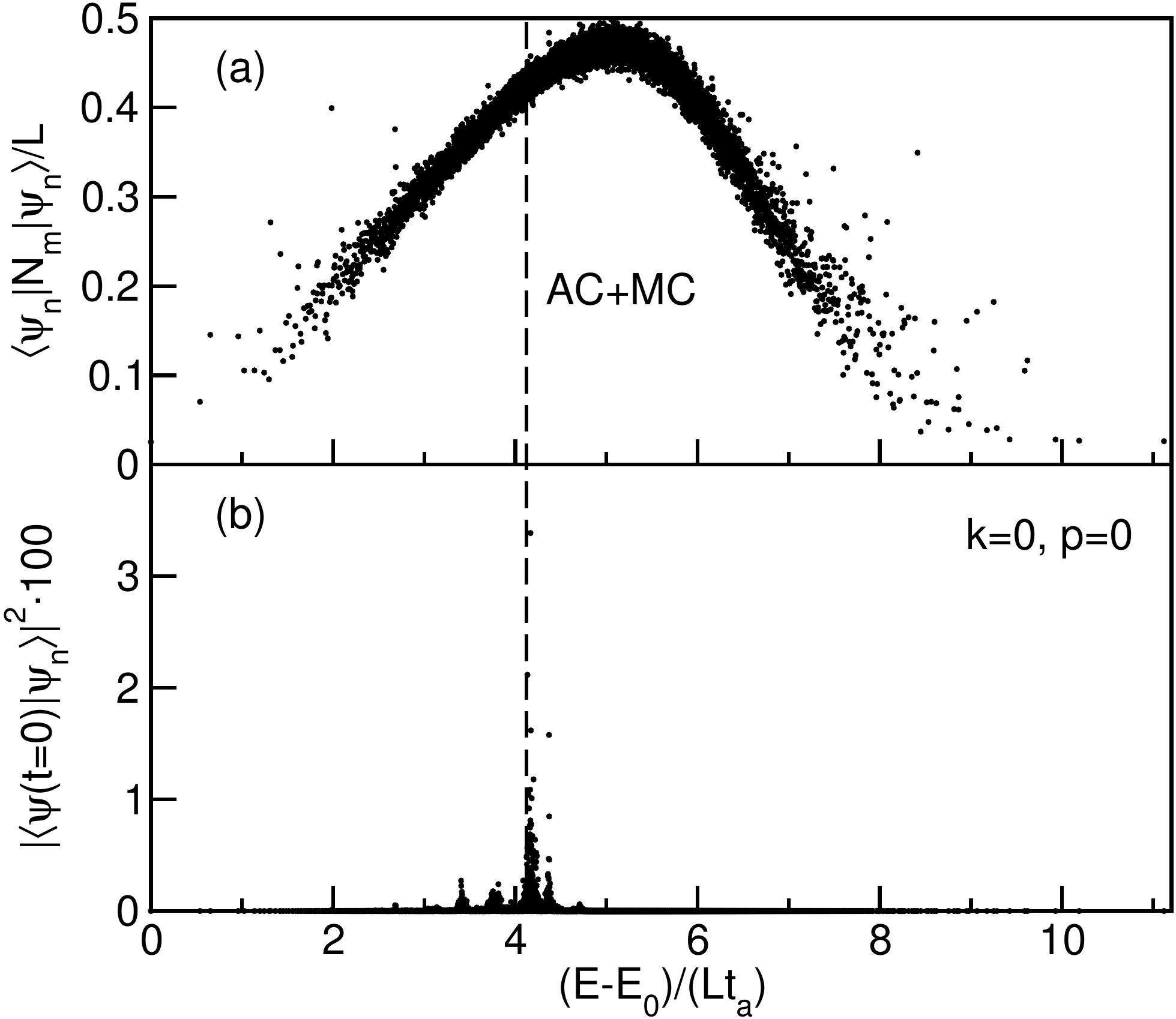}
\caption{(Color online)
(a) Distribution of postquench expectation values of the molecule particle number operator in the AC+MC phase along with (b) the overlaps of the initial state with the eigenstates in a system of size $L=6$.
The dashed line indicates the quench energy.
($\epsilon_{\rm m}= 2t_{\rm a}, t_{\rm a}/{U} = 3$). }\label{fig:specACMC}
\end{figure}

\begin{figure}[t]
\includegraphics[width=.96\columnwidth]{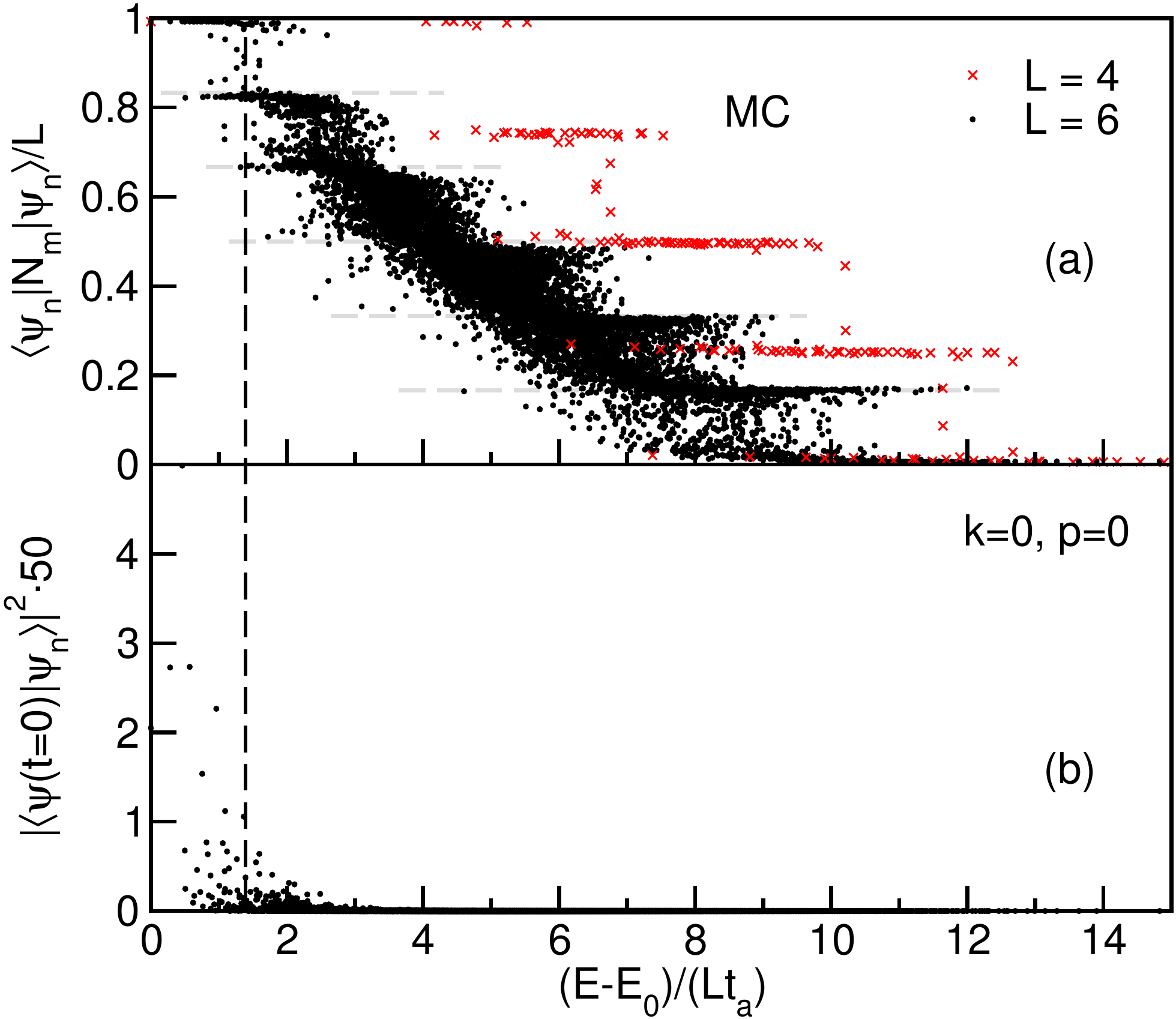}
\caption{(Color online)
(a) Distribution of postquench expectation values of the molecular particle number operator in the MC phase along with (b) the overlaps of the initial state with the eigenstates.
The dashed line indicates the quench energy. ($\epsilon_{\rm m}= 2t_{\rm a}, t_{\rm a}/{U} = 3$).
}
\label{fig:specMC}
\end{figure}

We now consider the distribution of postquench eigenstates expectation values of the molecular density as an example.
First, let us note that for our system,
 the initial-state overlap $|c_n|^2$ with the eigenstates of the postquench Hamiltonian is already a relatively narrow function of the energy on the system sizes considered
(see Figs.~\ref{fig:specACMC}(b) and Fig.~\ref{fig:specMC}(b)).

Figure~\ref{fig:specACMC}(a) shows the distribution of postquench eigenstate expectation values   for the quench from the MC to the AC+MC phase in a system of size $L=6$.
Already on such a small system, the DPQEV is a smooth and fairly sharp distribution for energies located in the bulk of the eigenspectrum.
By comparing to the  overlap of the initial state with the eigenstates of the postquench Hamiltonian
plotted in Fig.~\ref{fig:specACMC}(b), 
 we find that the initial state is very sharply peaked at an energy $(E-E_0)/L \approx 4 t_{\rm a}$ which is well inside the bulk of the eigenspectrum where the DPQEV is 
a sharp distribution, practically  depending only on energy.
We therefore conclude that in this case the ETH works (as expected for a generic quantum many-body system) 
and thus the system thermalizes, consistent with our numerical observations.

Figure~\ref{fig:specMC} shows the DPQEV for the quench from the MI to the MC phase for system sizes $L=4,6$.
The data for $L=4$ show plateaus at integer values of the number of molecules (similar to the double occupancy in the strongly interacting regime of the Fermi-Hubbard model \cite{bauer15}).
Increasing the system size introduces more eigenstates with intermediate (i.e., non-integer) molecular particle numbers but a general plateau structure can still be discerned.
The initial state overlaps $|c_n|^2$ are again a strongly peaked function of the energy.
Comparison to the DPQEV shows that it does not sample the bulk of the system but that this quench puts the system at the edge of the spectrum, where the
ETH is expected to work only for very large systems (see, e.g., \cite{rigol09,roux09,roux10,sorg14}).
Nevertheless, our numerical results indicate a reasonable agreement between the diagonal and thermal ensembles already on fairly small systems.

\subsection{Structure of optimal modes}
\label{sec:quenchstrucom}

\begin{figure}[t]
\includegraphics[width=.96\columnwidth]{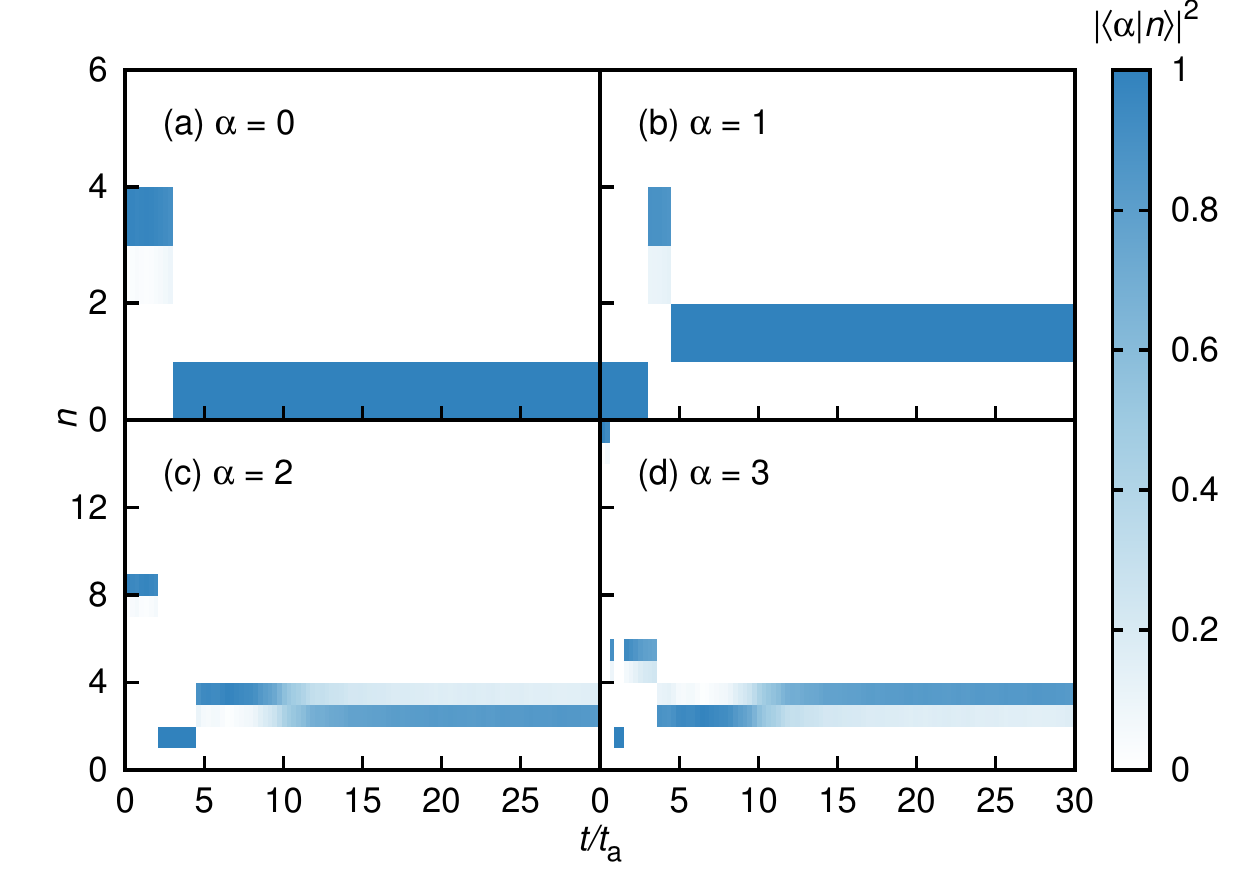}
\caption{(Color online)
(a)-(d) Evolution of the first four optimal modes $|\alpha \rangle $ in time along the trajectory from the MC ($\epsilon_{\rm m} = -6t_{\rm a}, \frac{t_{\rm a}}{U} = 3$) to the AC+MC phase ($\epsilon_{\rm m} = 2t_{\rm a}, \frac{t_{\rm a}}{U} = 3$).
The figure shows the weights $|\langle \alpha | n\rangle|^2$ of the bare local states $|n\rangle=|N_{\rm a}^{(1)}(n),N_{\rm m}^{(1)}(n)\rangle$ contributing to the optimal modes as a function of $t_{\rm a}/U$.
ED results for $L=8$.
The respective physical state corresponding to the index $n$ is defined in Tab.~\ref{tab:localbasis}.
}
\label{fig:modes_t_1}
\end{figure}

This section illustrates the dynamics of the optimal-mode spectra for the two quenches.
Results for the first quantum quench are shown in Fig.~\ref{fig:modes_t_1}(a)-(d).
For small values of $t/t_a$,  the mode spectra are the ones from the MI phase as shown in Fig.~\ref{fig:m_gs}.
Generally, the spectra change significantly as a function of time:
They start from states in the $N_{\rm T} = 0,2,4,6$ subspaces and change into states in the $N_{\rm T} = 0,1,2$ subspaces.
We can compare those to Fig.~\ref{fig:modes2} and find that after some time only the $\alpha = 1$ mode (Fig.~\ref{fig:modes_t_1}(b)) is the same as in the ground state deep in the AC+MC phase while the other three evolve to different structures.
Of course, this is no surprise as during the quench we pump energy into the system, leading to a final state that is generally not the post-quench ground state.
Apart from this we can read off that the contribution of atoms increases.
This is visible in the final distribution of the bare states:
The three most important bare modes in the steady state after the quench are the ones with $N_{\rm a}^{(1)} = 0,1,2$.
Also, we see in the third and fourth mode that the molecular contribution gets suppressed with time.
The high weight of the $N_{\rm T}^{(1)} = 0$ mode implies that there are increased fluctuations in local particle number in the steady state.

\begin{figure}[t]
\includegraphics[width=.96\columnwidth]{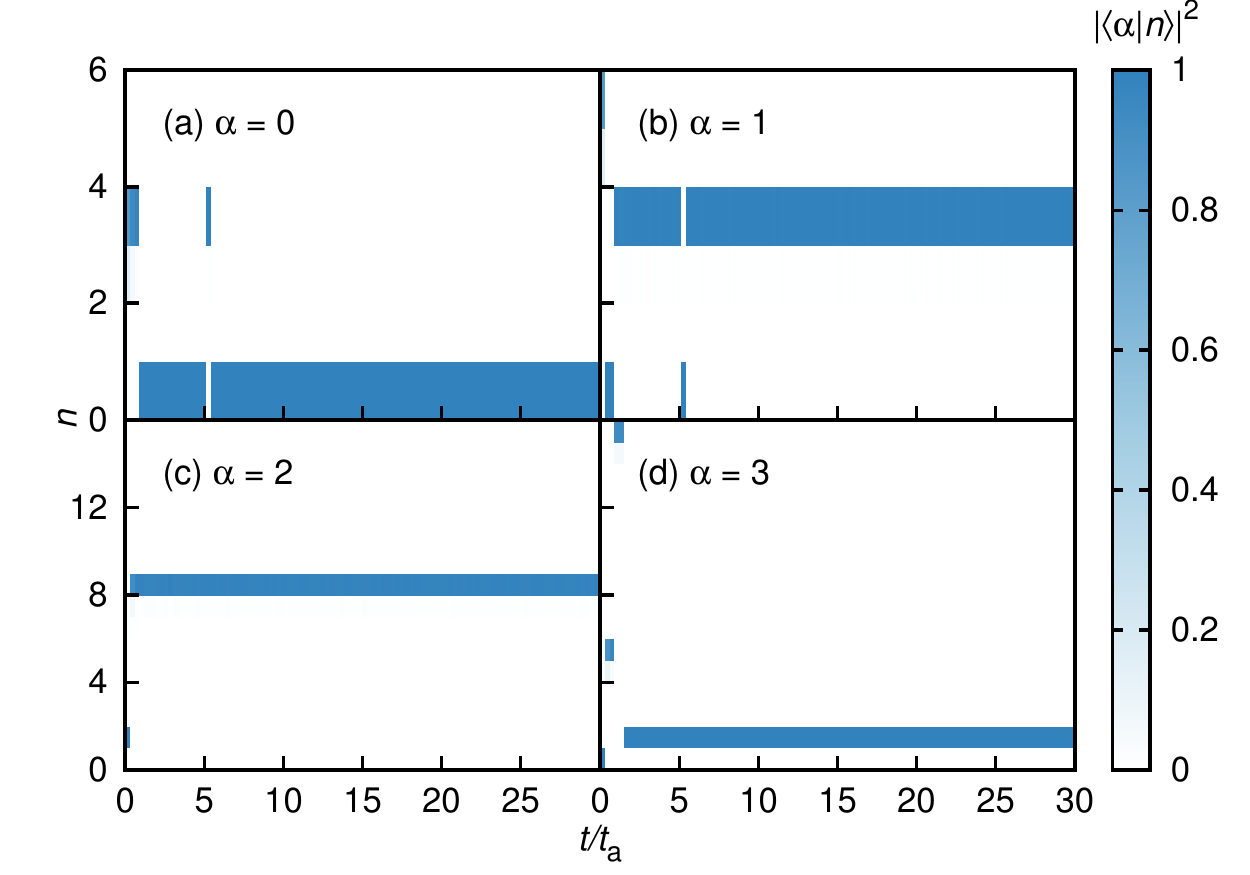}
\caption{(Color online)
(a)-(d) Evolution of the first four  optimal modes $|\alpha \rangle $ in time along the trajectory from the MI ($\epsilon_{\rm m} = -6t_{\rm a}, \frac{t_{\rm a}}{U} = 0.1$) to the MC phase ($\epsilon_{\rm m} = -6t_{\rm a}, \frac{t_{\rm a}}{U} = 3$).
The figure shows the weights $|\langle \alpha | n\rangle|^2$ of the bare local states $|n\rangle=|N_{\rm a}^{(1)}(n),N_{\rm m}^{(1)}(n)\rangle$ contributing to the optimal modes as a function of $t_{\rm a}/U$.
ED results for $L=8$.
The respective physical state corresponding to the index $n$ is defined in Tab.~\ref{tab:localbasis}.
}
\label{fig:modes_t_2}
\end{figure}

Results for the second quench are shown in Fig.~\ref{fig:modes_t_2}(a)-(d).
We see that the evolution to the final optimal-mode structure is very fast (takes less than $\sim 2t_{\rm a}$).
This is similar to the behavior of the quantities shown in Fig.~\ref{fig:nk_nm_svn_t_2}.
In Fig.~\ref{fig:nk_nm_svn_t_2} we see that the number of molecules is not greatly influenced by the quench.
The fact that the highest weighted bare mode is again the $N_{\rm T}^{(1)} = 0$ one suggests that fluctuations increase as time progresses.

Generally, the largest changes in the optimal-mode spectra happen over the same time window in which the local entropy $S_{\rm{vN}}^{(1)}$ varies significantly.
To find out if the optimal modes are thermal we compare the mode spectrum in the steady state with the optimal modes calculated in the canonical ensemble for both quenches.
In both cases (results not shown here) we find a strong similarity and therefore conclude that the optimal modes are thermal.

\section{Conclusion}
\label{sec:conclusion}
In summary, we analyzed the single-site reduced density matrix in the Bose-Bose resonance  model both in equilibrium and in quantum quenches.
As an example we considered the case of double filling $N_{T} = 2L$.
Since this model features two bosonic species, atoms and molecules, and since their individual particle numbers are not conserved, the local reduced density matrix is not diagonal in the basis of local bare modes (being eigenstates of both atomic and molecular particle number).
The analysis of the equilibrium properties shows that phase transitions can lead to features in the local von Neumann entropy.
For the phase transition between the MC and AC+MC phase one has to consider the first derivative of the local von Neumann entropy with respect to the detuning parameter.
At the boundary, this quantity shows a sharp maximum.
For the phase transition from the MI to the MC phase one can use the different $L$-dependence of the local von Neumann entropy on the MI and MC sides of the phase boundary.
In the MI phase, the local von Neumann entropy saturates to a finite value from below while in the MC phase we show that it saturates to a finite value from above.
The point at which this monotony behavior in the system-size dependence changes is close to the known value for this phase transition.

We further studied the optimal modes and their weights as a function of the control parameters for both trajectories through the phase diagram.
These quantities are shown to be different when one considers the system at a point deep in either one of the three phases.
Monitoring the change of the first few most important modes as a function of the control parameters along the two trajectories we conclude that they also reflect the phase transition.

Finally, we performed two quantum quenches along the two trajectories where we start from the ground state in one phase and quench the system over to the final parameters.
For those quenches we study the fraction of atoms and molecules which are at quasimomentum $k=0$, the density of molecules and the local von Neumann entropy as a function of time.
We also compute the diagonal and canonical ensemble averages for the molecular density and find that both agree quite well with the steady-state value.
The good agreement with the canonical expectation value  is, for the first quantum quench, explained by the sharply peaked initial state and the sharp
distribution of postquench eigenstate expectation values (and therefore the realization that the conditions for the eigenstate thermalization hypothesis to apply are fulfilled, already 
on small 
systems).

We finally considered the optimal-mode spectra as a function of time and 
observe that their steady-state structure is clearly different from the structure in  both the initial state and in the ground state at the postquench parameters.
Most importantly, the comparison of the steady-state values of the single-site entanglement entropy and the optimal modes to the canonical ensemble shows that the 
single-site reduced density matrix is thermal in the steady state.

The decomposition of the local reduced density matrix into weights and optimal modes does not only give physical insight but can also be used to design numerical methods \cite{jeckelmann98,guo12,brockt15}, following the ideas of \cite{zhang99},
suggesting to set up an effective basis using the eigenstates of single-site reduced density matrices plus truncation in their spectrum.
In principle, this method can be used to greatly reduce the local state space in a controlled way.
This has been shown to work for the ground state of the Holstein model using ED \cite{jeckelmann98}, a matrix-product state method applied to 
the spin-boson model \cite{guo12}, and more 
recently, also for the  time evolution in electron-phonon problems using the time evolving block decimation algorithm \cite{brockt15}.
The application of such ideas to ground-state DMRG algorithms for electron-phonon systems is an open problem.
Considering the behavior of the single-site entanglement entropy and of the optimal modes, 
the Hubbard-Holstein model is an interesting candidate for further studies,
since the  structure of the optimal modes can be richer than in the system considered here.
A further sophistication of the model would be to allow for a dispersion of phonons.

We acknowledge stimulating and helpful discussions with Claudius Hubig, Eric Jeckelmann, Jacopo Nespolo and Lev Vidmar.
F.D. and F.H.-M. acknowledge support from the DFG (Deutsche Forschungsgemeinschaft) through Grant No. HE 5242/3-1 in the Research Unit Advanced Computational Methods for Strongly Correlated Quantum Systems (FOR 1807).


\appendix

\section{Local von Neumann entropy in the SF case}
\label{app:perttheory}

This section details the calculation of the behavior of the local von Neumann entropy as a function of system size for the case of the non-interacting Bose-Hubbard model.
For this system the weights of the local reduced density matrix can be calculated exactly \cite{ding09}
\begin{align}
\lambda_l = L^{-N} \binom{N}{l} (L-1)^{(N-l)}.
\end{align}
For unit filling we get
\begin{align}
\lambda_l = \frac{(L-1)^{L-l}}{L^{L}} \binom{N}{l}.
\end{align}
In order to go on we use an upper bound for the binomial factor $\binom{L}{l} \approx \frac{L^l}{l!}$
\begin{align}
\lambda_l = \left( \frac{L-1}{L} \right)^{L-l} \frac{1}{l!}.
\end{align}
We define
\begin{align}
F_l(L) =& \left( \frac{L-1}{L} \right)^{L-l} \\
\frac{\partial}{\partial L} F_l(L) =& \left( \ln \left(\frac{L-1}{L}\right)+\frac{1}{L-1}-\frac{l}{L(L-1)} \right) F_l(L).
\end{align}
We can now derive the behavior of the local von Neumann entropy
\begin{align}
S_{\rm vN}^{(1)}(L) =& -\sum\limits_l \frac{F_l(L)}{l!} \ln\left( \frac{F_l(L)}{l!} \right) \label{eq:svnapp} \\
\frac{\partial}{\partial L} S_{\rm vN}^{(1)}(L) =& \sum\limits_l \frac{F_l'(L)}{l!} (\ln(l!) - ln(F_l(L)) - 1).
\end{align}
We are interested in large systems so we go to the asymptotic limit
\begin{align}
\lim_{L \rightarrow \infty} F_l(L) \approx& \frac{1}{\rm e} + \frac{l-1/2}{{\rm e} L} + \frac{12 l^2-5}{24 {\rm e} L^2} \label{eq:fl} \\
\lim_{L \rightarrow \infty} F_l'(L) \approx& \frac{1}{\rm e} (\frac{1}{2}-l)\frac{1}{L^2}, \label{eq:dfl}
\end{align}
where all terms were kept up to second order in $1/L$.
The derivative of the local von Neumann entropy then becomes
\begin{align}
\lim_{L \rightarrow \infty} {S_{\rm vN}^{(1)}}'(L) \approx& \sum\limits_l \frac{{F_l}'(L)}{l!} \ln(l!).
\end{align}
Inspecting $F_l'(L)$ we see that this quantity is always negative except for $l = 0$.
Since the $\ln(l!)$ term kills all terms in the sum with $l<2$ we see that the derivative of the local von Neumann entropy with respect to system size $L$ is always negative.
Plugging in Eqs.~\eqref{eq:fl} and \eqref{eq:dfl} into Eq.~\eqref{eq:svnapp} we see that $S_{\rm vN}^{(1)}$ approaches its asymptotic value from above with a $1/L$ correction.

\bibliography{references}
\end{document}